# Destabilization of free convection by weak rotation

A. Yu. Gelfgat

School of Mechanical Engineering, Faculty of Engineering, Tel-Aviv University, 69978, Tel-Aviv, Israel

**Abstract**

This study offers an explanation of a recently observed effect of destabilization of free convective flows by weak rotation. After studying several models where flows are driven by a simultaneous action of convection and rotation, it is concluded that the destabilization is observed in the cases where centrifugal force acts against main convective circulation. At relatively low Prandtl numbers this counter action can split the main vortex into two counter rotating vortices, where the interaction leads to instability. At larger Prandtl numbers, the counter action of the centrifugal force steepens an unstable thermal stratification, which triggers Rayleigh-Bénard instability mechanism. Both cases can be enhanced by advection of azimuthal velocity disturbances towards the axis, where they grow and excite perturbations of the radial velocity. The effect was studied considering a combined convective/rotating flow in a cylinder with a rotating lid and a parabolic temperature profile at the sidewall. Next, explanations of the destabilization effect for rotating magnetic field driven flow and melt flow in a Czochralski crystal growth model were derived.

## 1. Introduction

This study is devoted to an effect of destabilization of axisymmetric natural convection flows by a weak superimposed non-uniform rotation. In classical models, such as a rotating infinite layer (see, e.g., Chandrasekhar, 1961; Koschmieder,1993; Fernando & Smith IV, 2001; Kloosterziel & Carnevale, 2003; Lewis, 2010) or rotating cylinders and annuli (see, e.g., Lucas et al., 1983; Goldstein at al., 1993, Lopez & Marquez, 2009; Rubio et al., 2010), increasing rotation usually leads to stabilization of the flow, i.e., to the growth of the critical Rayleigh number or other critical parameters describing magnitude of the buoyancy force. We do not review here numerous studies of the two above mentioned models, but address the reader to references in the cited papers. It seems that there exists a common agreement that the effect of rotation on convective instabilities is mainly stabilizing. However, as is shown in this paper, such a generalization is wrong. The two above models consider classical Rayleigh-Bénard problem of stability of purely conducting quiescent states, while in most practically important cases, the buoyancy force is non-potential, so that natural convection flow always exists. Brummel et al. (2000) argued that non-uniform rotation can destabilize even stably stratified non-isothermal flow. Reviewing several studies devoted to instabilities driven by the simultaneous effect of convection and rotation, Koschmieder (1993) noted that when the rotation affects the base flow, its effect on flow instability becomes very complicated. We can add that non-uniform rotation caused by a rotating boundary or an external force complicates the stability properties of flows even more. This was observed, for example, in non-isothermal Taylor-Couette flow (Ali & Weidman, 1990; Ali & McFadden, 2000), where circular Couette flow and an infinite convective loop were superimposed in the base flow state.

This study is motivated by several recent observations of destabilization of convective flow by rotation in models of Czochralski bulk crystal growth process (see, e.g., Müller, 2007). These models consider melt flow in a cylindrical crucible with a heated bottom and sidewall, cooled by a rotating cold crystal pulled out from the upper free surface. In laboratory flow models, the crystal is usually replaced by a cooled cylindrical dummy whose lower surface touches the free surface of working liquid. Rotation and lower temperature of the dummy mimic the effect of the crystal in a technological setup (Hintz et al., 2001; Schwabe et al, 2004; Teitel et al., 2008). The flow is driven by buoyancy, rotation of the crystal and thermocapillary force acting along the free surface. Stability studies of these flows focus on parameters at which steady – oscillatory flow transition takes place. Different examples can be found in Gelfgat (2008) and references therein. Figure 1 shows two examples of stability diagrams, in which critical temperature difference $\Delta T_{cr}$, to which both Grashof and Marangoni numbers are proportional, is plotted versus the Reynolds number defined by the angular velocity of the dummy rotation and the crucible radius. The system geometry and an example of the flow pattern are shown in the insert. Further details can be found in Teitel *et al.* (2008), Crnogorac *et al.* (2008), and Gelfgat (2008). Both examples relate to similar experiments with different working liquids with the Prandtl number $Pr=9.2$ in the case (a) of Schwabe *et al.* (2004) and $Pr=23.9$ in the case (b) of Teitel *et al.* (2008). In both examples computations predict a steep decrease of $\Delta T_{cr}$ with a slow increase of the rotational Reynolds number up to $Re=100$, which corresponds to the rotation of crystal with the angular viscosity smaller than 0.1 and 0.5 rad/s (1 and 5 revolutions per minute) in the cases (a) and (b), respectively. Note that depending on the aspect ratio, isothermal swirling flow in a cylinder with a rotating lid becomes unstable for the Reynolds number between 2000 and 3000 (Gelfgat *et al.*, 2001), so that the destabilization observed in Fig. 1 cannot be addressed to a rotation-induced instability. Note also that in spite of the critical temperature difference $\Delta T_{cr}$ decreases in more than an order of magnitude, it never reaches zero. Therefore, it is an interaction of all the driving forces that makes the flow significantly lesser stable.

At the present time there is no thorough experimental evidence of the destabilization predicted by the computational modelling. Some qualitative evidence of this phenomenon can be found in the observations of Teitel *et al.* (2008), which also support our computations (Fig. 1b), as well as in Kakimoto *et al.* (1990), Munakata & Tanasawa (1990), Ozoe *et al.* (1991), Kishida *et al.* (1993), Seidl *et al.* (1994), and Suzuki (2004). Some independent numerical results exhibiting the Czochralski flow destabilization with increasing rotation can be found in Sung *et al.* (1995), Akamatsu *et al.* (1997), Zeng *et al.* (2003) and Banerjee & Muralidhar (2006). The fluid Prandtl number in these works varies from $Pr = 0.011$ in Kishida *et al.* (1993) to $Pr = 23.9$ in Teitel *et al.* (2008), and even $Pr \approx 4600$ in Ozoe *et al.* (1991) and Sung *et al.* (1995), so that the destabilization phenomenon can be expected and, as it is shown below, appears in flows with significantly different Prandtl numbers.

Since the Czochralski flow model is rather complicated, in the present study we are looking for a simpler characteristic model exhibiting similar destabilization and intend to study the latter to get more physical insight in the phenomenon. Considering several examples of flows driven by convection and rotation we show that the simplest model exhibiting similar destabilization is a combination of two well-studied cases: convective flow in a vertical cylinder with a parabolic temperature profile at the sidewall, and swirling flow in a cylinder with a rotating lid. The three-dimensional stability of the first one was studied in Gelfgat *et al.* (2000) and of the second one in Gelfgat *et al.* (2001). Examining the flow and leading disturbance patterns we arrive at the conclusion that destabilization is caused by a counter action of the centrifugal force that tends to slow down the main convective vortex. This counter action leads either to the appearance of a new vortical structure or to a steeper temperature gradient along the axis. Consequently, the following destabilization is connected either to an interaction between counter rotating vortices, or to the Rayleigh-Bénard instability developing below the cold upper boundary. An additional destabilization mechanism is connected to the advection of perturbations of azimuthal velocity towards the axis, where they grow due to conservation of the angular momentum, which causes a growth of the radial velocity disturbances. It is emphasized that we focus here only on the phenomenon of destabilization and do not pretend to supply a comprehensive description of a variety of instabilities that take place in convective, rotating, as well as combined convective/rotating flows, whose variety hardly can be described within a single journal paper. On the other hand, to show that the described destabilization phenomena can be expected in other convective/rotating flows, we add an example in which buoyancy force interacts with a rotating magnetic field driving effect.

## 2. Problems formulation and numerical technique

We study three-dimensional instabilities of axisymmetric non-isothermal base flows. The full three-dimensional problem is described by the Boussinesq equations in cylindrical coordinates. A Boussinesq fluid with density $\rho^*$, kinematic viscosity $\nu^*$ and thermal diffusivity $\chi^*$ in an axisymmetric region $0 \le r \le R^*$, $0 \le z \le H^*$ is considered. The polar axis is assumed to be parallel to the gravity force. The flow is described by the momentum, continuity and energy equations in cylindrical coordinates $(r^*, z^*)$. To render the equations dimensionless, we use the scales $R^*$, $R^{*2}/\nu^*$, $\nu^*/R^*$, $\rho^*(\nu^*/R^*)^2$ for length, time, velocity and pressure respectively. The temperature is rendered dimensionless by the relation $T = (T^* - T^*_{cold})/(T^*_{hot} - T^*_{cold})$, where $T^*_{hot}$ and $T^*_{cold}$ are the maximal and minimal temperatures at the boundaries of the flow region. The set of Boussinesq equations for the non-dimensional velocity $\mathbf{v} = \{v_r, v_\theta, v_z\}$, temperature $T$ and pressure $p$ in the domain $0 \le r \le 1$, $0 \le z \le A$ reads

$$\frac{\partial \mathbf{v}}{\partial t} + (\mathbf{v} \cdot \nabla)\mathbf{v} = -\nabla p + \Delta \mathbf{v} + GrT\mathbf{e}_z + Taf(r,z)\mathbf{e}_\theta \qquad (1)$$

$$\frac{\partial T}{\partial t} + (\mathbf{v} \cdot \nabla) T = \frac{1}{Pr} \Delta T, \qquad \nabla \cdot \mathbf{v} = 0 \tag{2,3}$$

Here $A = H^*/R^*$ is the aspect ratio, $Gr = g^* \beta^* (T^*_{hot} - T^*_{cold}) R^{*3}/\nu^{*2}$ the Grashof number, $Pr = \nu^*/\chi^*$ the Prandtl number, $g^*$ gravity acceleration, $\beta^*$ the thermal expansion coefficient, and $\mathbf{e}_z$ the unit vector in the $z$-direction. The last term in Eq. (1) describes time-averaged azimuthal force resulting from an externally applied rotating magnetic field (RMF), which is considered as a possible source of rotational motion in one of the examples below. This force is described by an analytical function (Gorbachev *et al.*, 1974; Grants & Gerbeth, 2001)

$$f(r,z) = r - 2\sum_{k=0}^{\infty} \frac{J_1(\gamma_k r) \cosh\left(\gamma_k \frac{2z - A}{2}\right)}{(\gamma_k^2 - 1) J_1(\gamma_k) \cosh\left(\gamma_k \frac{A}{2}\right)} \tag{4}$$

where $J_1$ is the Bessel functions and $\gamma_k$ are eigenvalues of the problem $J_1'(\gamma_k) = 0$. The non-dimensional parameter that defines the force magnitude is the magnetic Taylor number $Ta = \omega^* \sigma^* B_0^{*2} R^{*2}/(2\rho\nu^{*2})$, where $B_0^*$ and $\omega^*$ are the magnitude and rotational frequency of the magnetic field, and $\sigma^*$ is the electric conductivity of the fluid. Additionally, if the top or bottom of the cylinder rotate with an angular velocity $\Omega$, we define the rotational Reynolds number as $Re = \Omega R^{*2}/\nu$. Since the models considered below do not include background uniform rotation, we do not include centrifugal buoyancy effects in our formulation (see, e.g., Lopez & Marques, 2009; Rubio et al., 2010). The boundary conditions will be specified below, separately for each problem.

We study instability of steady axisymmetric flows $\{\mathbf{V}, P, T\}, \mathbf{V} = (U, V, W)$ with respect to infinitesimally small three-dimensional disturbances, which are decomposed into a Fourier series in the azimuthal direction and are represented as $\sum_{m=-\infty}^{m=+\infty} \{\tilde{\mathbf{v}}_m, \tilde{p}_m, \tilde{T}_m\}\, exp(\lambda t + m\theta)$, where $\lambda$ is a complex amplification rate, $\tilde{\mathbf{v}} = (\tilde{u}, \tilde{v}, \tilde{w})$, $\tilde{p}$ and $\tilde{T}$ are perturbation of the velocity, pressure, and temperature, respectively. The subscript $m$ denotes the $m$-th Fourier mode of a corresponding function. It is well-known that the linear stability problem separates for each $m$, which is an integer azimuthal wavenumber. Therefore, after an axisymmetric steady state is computed, solution of the stability problem is reduced to a series of 2D-like generalized eigenvalue problems defined for the eigenvalues $\lambda$ separately at different azimuthal wavenumbers $m$ (see, e.g., Gelfgat, 2007). The steady flow is unstable when at least one $\lambda$ exists with a positive real part. The eigenvalue with the largest real part is called leading and parameters at which the leading eigenvalue crosses the imaginary axis are called critical.

To calculate steady states of Eqs. (1)-(3) and to study their linear stability with respect to three-dimensional infinitesimal disturbances, we use the finite volume discretization and the technique described and verified in Gelfgat (2007). The test calculations performed there showed that to keep critical parameters within 1% accuracy, one needs to apply at least 100 grid points in the shortest spatial direction. Note that our stability results for isothermal flow in a cylinder with a rotating top were successfully compared with several independent computations (Gelfgat, 2002) and were validated experimentally by Sørensen et al. (2006). The results for stability of convective flow in a side-heated cylinder are also well-compared with the independent result of Gemeny et al. (2007). Together with the convergence studies (Gelfgat, 2007) these make us confident in the accuracy of the results reported below. In the following calculations, the size of the stretched finite volume grid varies from $N_r = 100$ to 300 points in the radial direction. The grid size in the axial direction is taken as $N_z = AN_r$. The grid size is chosen to ensure convergence to at least three correct decimal places in the calculated critical parameters.

## 3. Instability of axisymmetric flows driven by buoyancy convection and rotation: examples

To explain the destabilization described in the Introduction, we seek a simple model flow, which exhibits similar phenomenon, i.e., destabilization of convective circulation by slow rotation that takes place in a wide range of the Prandtl numbers. It can be easily checked that the steep destabilization of Czochralski flow persists if the thermocapillary force is set to zero. Therefore, thermocapillarity can be excluded from a qualitative model we are trying to find. We also exclude the classical Rayleigh-Bénard configuration of the cylinder with an isothermal top and bottom, since this problem can lead to multiple secondary states (Boronska & Tuckerman, 2010), so that each distinct steady state branch will be affected differently by the same rotational forcing. In this study we prefer to refrain from configurations allowing for multiple steady states.  Figures 2-4 represent several characteristic examples that illustrate the existence or absence of the destabilization and help us to focus on a model that seems to be the most appropriate for our purposes. Stability diagrams shown in these figures for different thermal boundary conditions and different Prandtl numbers are supplied by examples of streamlines and isotherms.

It should be noticed that almost all instabilities observed are oscillatory, i.e., the corresponding leading eigenvalue has a non-zero imaginary part. The imaginary part estimates the frequency of appearing oscillations. The latter is not shown on separate graphs because it does not relate directly to the destabilization effect discussed. It is quite obvious that if a perturbation is not axisymmetric ($m \neq 0$) and the base flow has a non-zero azimuthal component, then the non-axisymmetric perturbation pattern is advected around the axis, which makes steady three-dimensional instability impossible for $Re \neq 0$. Therefore, steady bifurcation can appear either for the axisymmetric Fourier mode $m=0$ or for non-rotating basic flows at $Re=0$. Among the results shown in Figs. 2-4 only four points correspond to the steady instability: in Fig. 2a point $Re_{top}$=0, $Pr$=0.7; in Fig. 3c two points at $Re_{bottom}$=0; and in Fig. 4c the point $Re_{top}$=0. All the other points on the neutral stability curves correspond to the oscillatory instability.

In all the cases we impose no-slip boundary conditions on all the boundaries. Assuming that either the top or bottom of the cylinder rotates with a constant angular velocity, $\Omega_{top}$ or $\Omega_{bottom}$, respectively, the boundary conditions read

$$\text{at } r = 1: \qquad v_r = v_\theta = v_z = 0 \tag{5}$$

$$\text{at } z = 0: \qquad v_r = v_z = 0, \quad v_\theta = Re_{bottom} r \tag{6}$$

$$\text{at } z = A: \qquad v_r = v_z = 0, \quad v_\theta = Re_{top} r \tag{7}$$

The temperature boundary conditions will be defined separately for each problem. Additionally, they are summarized in Table 1 together with parameters at which the flow patterns are reported in Figs. 2-4. The critical azimuthal wavenumbers $m_{cr}$ as well as values of the Prandtl number and the aspect ratio, are shown in the figures.

Figures 2a and 2b show stability diagrams for vertical cylinders non-uniformly heated from below: by a parabolic radial temperature profile (Fig. 2a)

$$T(r, z = 0) = 1 - r^2, \quad T(r, z = A) = 0, \quad \left(\frac{\partial T}{\partial r}\right)_{r=1} = 0 \tag{8}$$

and by heating of a half-radius inner part of the boundary (Fig. 2b)

$$T(r \le 0.5, z = 0) = 1, \quad T(r > 0.5, z = 0) = 0, \quad T(r, z = A) = 0, \quad \left(\frac{\partial T}{\partial r}\right)_{r=1} = 0\ . \tag{9}$$

In both cases the convective flow is affected by rotation of the upper boundary ($Re_{bottom} = 0$). In both cases the fluid driven by the buoyancy force rises along the axis and descends along the sidewall. The rotation of the top creates a centrifugal force that pushes the fluid from the center to the sidewall along the upper boundary, and therefore intensifies the convective circulation. Depending on the Prandtl number, we observe either existence or absence of destabilization of convection by rotation. Surprisingly, in the case of Fig. 2a the destabilization exists for larger Prandtl number, $Pr=7$, while in the case of Fig. 2b – for the smaller Prandtl number, $Pr=0.7$. In these cases the critical Grashof number can be reduced by approximately four to five times. Thus, these two examples show that a significant destabilization can be observed at certain conditions, but its existence depends on the ratio of momentum and heat dissipation, i.e., the Prandtl number, so that no general conclusion can be drawn and each particular case should be studied separately. We conclude that these two models are not the characteristic model we are looking for.

Figure 2c presents another example where the convective circulation is created by a parabolic temperature profile at the cylinder sidewall, while the top and bottom remain at a low constant temperature

$$T(r,z=0) = T(r,z=A) = 0, \quad T(r=1,z) = 4z(1-z). \tag{10}$$

In this case the fluid ascends along the sidewall and descends along the cylinder axis. Stability of this convective flow was studied by Gelfgat et al. (2000). For the purposes of this study we affect the convective circulation by an azimuthal RMF force (4) keeping $Re_{bottom} = Re_{top} = 0$. Three-dimensional instability of the isothermal RMF-driven rotating flow was studied by Grants & Gerbeth (2002), who showed that the critical magnetic Taylor number exceeds the value of $Ta=10^5$. In our example the Taylor number does not exceed $10^4$, so that the RMF-driven flow instabilities do not develop. Again, we observe that a relatively weak RMF force with $Ta \approx 10^3$ leads to a decrease of the critical Grashof number by more than five times. At larger Taylor numbers, beyond the minimum, $Gr_{cr}$ monotonically grows with the increase of $Ta$. Note that the observed destabilization takes place with the convective circulation rotating in the opposite direction compared to Figs. 2a and 2b. Moreover, unlike the examples in Figs. 2a and 2b, the RMF force drives rotation of the whole fluid volume, so that the angular velocity grows from the no-slip top and bottom towards the midplane $z = 0.5$, where the centrifugal force attains its maximal value. Thus, without convection or at larger $Ta$ the meridional flow consists of two circulations (Grants & Gerbeth, 2001); the insert in Fig. 2c corresponding to $Ta$=$10^4$ shows the beginning of the second circulation development. This example shows again that destabilization of convective circulation by a weak non-uniform rotation is a rather common phenomenon and is not necessarily connected with a rotation of one of boundaries. Formally, we can consider the RMF-driven flow also at the larger Prandtl number. However, it would be meaningless because the Prandtl number of electrically conducting fluids does not exceed 0.1.

Figures 3 and 4 present stability diagrams for the convective flow resulting from the same sidewall parabolic heating, as in Eq. (10) and Fig. 2c and affected by rotation of either the bottom (Fig. 3) or the top (Fig. 4). Frames (a), (b), and (c) correspond to different Prandtl numbers 0.015, 0.7 and 7, respectively. In addition, Fig. 4a contains the neutral curve for a limit case $Pr = 0$, that corresponds to a very large thermal diffusivity. Note that at smaller Prandtl numbers, that can be interpreted as fluids with smaller viscosity, we consider larger intervals of the Reynolds number to cover all the values of our interest.

It can be stated immediately that no significant destabilization is observed when the convective flow is affected by rotation of the bottom boundary (Fig. 3). A slight destabilization for $Pr = 0.015$ cannot be compared to what was observed in Figs. 1 and 2. In the case of a rotating bottom the centrifugal force drives the flow from the axis towards the sidewall along the bottom and therefore intensifies the convective circulation. In spite of that at $Pr = 0.7$ and 7 we observe stabilization of the flow. At larger Prandtl numbers (Fig. 3c, $Pr = 7$) the instability sets in as a spoke pattern, which is illustrated by the temperature perturbation isolines plotted in an axial cross-section just below the upper boundary. This type of instability was observed experimentally and numerically by Gelfgat *et al.* (1999), where we argued that it is caused by the Rayleigh-Bénard mechanism developing in the upper unstably stratified flow region. Some more details on this instability were reported later by Szmyd *et al.* (2002). It also follows from Fig. 3b that weak bottom rotation can trigger the spoke pattern instability, which is a result of intensifying heat convection and thinning of the thermal boundary layer near the upper boundary.

A combination of the parabolic sidewall heating and rotation of the top yields the desirable effect (Fig. 4): independent on the Prandtl number, a relatively weak rotation destabilizes the convective circulation so, that the critical Grashof number decreases in about an order of magnitude. In this case the centrifugal force drives the flow from the axis to the sidewall along the upper boundary, so that it counteracts the convective circulation. As a result, at small *Re* we observe a retardation of the flow near the top, and with further increase of the Reynolds number an opposite circulation develops. This configuration is similar to the Czochralski flow model (Fig. 1), motivating our study by exhibiting the steep destabilization of convection by rotation, as well as by relative action of the buoyancy and centrifugal forces. The destabilization is observed in a wide range of the Prandtl numbers. We consider this case as the characteristic one and will study it in detail below.

Another similarity between the Czochralski configuration and the one chosen as "characteristic" can be observed in the pattern of supercritical oscillatory flow (Fig. 5). Shlieren visualizations made by Teitel *et al.* (2008) revealed so-called "cold plumes" instability, where cold fluid accumulates near the upper cold boundary and then descends towards the bottom. A similar instability was observed by Munakata & Tanasawa (1990) but for much larger Prandtl number, *Pr*=1000 (see Fig. 2 of Teitel *et al.*, 2008 and Figs. 5 and 6 of Munakata & Tanasawa, 1990, and multimedia files related to our Fig. 5). A time-dependent calculation for our characteristic model (Fig. 5b) also reveals a cold plume descending from the upper towards the lower boundary. Note that an attempt to reveal cold plumes from linear stability analysis only is not always successful. This is shown in Fig. 5a where supercritical oscillatory flow is approximated by a superposition of the base flow and the most unstable perturbation. The perturbation amplitude at a small supercriticality can be calculated as in Gelfgat et al. (1996). For present illustration purposes the supercriticality was taken small enough to keep the temperature in the interval $0 \le T \le 1$. The superposition also reveals plumes of cold fluid descending along the axis, however it does not reveal the cold thermal shapes observed in Fig. 5b and experiments of Teitel *et al.*, 2008 (see also multimedia files related to Fig. 5). This example shows, in particular, how cautious one should be when the linear stability results are extrapolated into a non-linear regime, and vice versa when linear instability arguments are derived from fully nonlinear observations.

## 4. More details on destabilization mechanisms

### *4.1. The characteristic model*

In the following study of mechanisms that cause destabilization of convective flow, we take into account that in all the cases reported in Fig. 4 the rotation is slow. In fact, the critical Reynolds number of the corresponding swirling isothermal flow is beyond 2000 (Gelfgat et al., 2001), so that one cannot expect appearance of any purely rotation induced instabilities for $Re \le 600$. With this in mind, we focus on three general possibilities. The destabilization can be caused either (i) by alteration of a leading disturbance by rotation making it unstable, (ii) by alteration of the base meridional flow pattern that leads to a destabilization of one of the leading eigenmodes, and (iii) by a new eigenmode whose appearance is caused by the base rotational motion. Note that in the option (i) we assume that a decaying disturbance existing without rotation is thereby destabilized, while alteration of the base meridional flow seems to be negligible. In the option (ii), we expect to notice some considerable changes in the meridional flow and assume that if a similar base flow existed without rotation it would be destabilized by a similar meridional disturbance. Finally, in the option (iii) we expect to observe an eigenmode that does not exist in the non-rotational case. Obviously, these possibilities relate to each other and can exhibit a variety of changes in the patterns of the base flows and perturbations. Moreover, we can expect a simultaneous observation of different possibilities, which would result in their interaction.

To illustrate some of the above destabilization possibilities, we choose several characteristic points on the neutral stability curves of Fig. 4 and continue the study in the following way. First, we exclude from the linearized stability problem the terms containing the base azimuthal velocity component $V$, and observe changes in the critical Grashof numbers (Table 2). A significant increase in the critical value would indicate on the term or terms contributing to the destabilization. Additionally we consider the terms $\boldsymbol{e}_{\theta} U\, \partial(r\tilde{v})/\partial r$ and $\boldsymbol{e}_{\theta} W\, \partial\tilde{v}/\partial z$ that describe advection of the azimuthal velocity perturbation by the meridional flow. A significant stabilization observed when these terms are switched off would indicate on the importance of the azimuthal velocity perturbation for the whole instability mechanism. Second, we plot the patterns of base flows at the same Grashof number with and without rotation to visualize alterations the rotation produces. Third, we plot the absolute values of the leading eigenvector of the linearized stability problem, which describe the distribution of oscillation amplitudes in the meridional plane. To simplify further explanations, the perturbation patterns are plotted together with the base flow streamlines. Together with the two previous steps, it allows us to derive some conclusions about the destabilization mechanisms. All the cases described and the conclusions derived are summarized in Table 3.

We start the discussion from the case *Pr*=0 (Fig. 4a), for which temperature perturbations do not play any role and the instability is of purely hydrodynamic nature. We consider point A on the corresponding neutral stability curve in Fig. 4a, which is located on a steeply decaying branch. Observations of switching on and off different terms of the linearized problem (Table 2) can be summarized as follows. Stabilization is observed when the terms containing either perturbation of the azimuthal velocity $\tilde{v}$ (case 14) or the base azimuthal velocity component (cases 5,7,11, 16,17), or both (cases 9,12,15) are switched off. The largest stabilization is observed in the cases 12 and 15 when all or almost all the convective terms of the linearized azimuthal equation are switched off. This shows that the destabilization is caused by advection of the base rotation, as well as its disturbance. There exists also a weaker stabilization connected with the switching of the terms related to the meridional flow (cases 16-18): switching of one or both of the terms containing the base and perturbed azimuthal velocity components $im\tilde{u}V\boldsymbol{e}_r/r$ and $2\tilde{v}V\boldsymbol{e}_r/r$ also leads to almost a doubling of $Gr_{cr}$. For a more detailed explanation, we refer to Fig. 6 where flow and perturbation patterns are presented.

In the upper part of the cylinder, where the radial velocity component is negative, the meridional flow advects the azimuthal momentum from the sidewall towards the axis (Fig. 6a). The rotational velocity in this region is maximal near the sidewall, and with the decrease of $r$ attains a local minimum and a local maximum reaching the zero value at the axis (Fig. 6e). Since the angular momentum tends to be preserved, the advection of rotation from the sidewall towards the axis necessarily leads to the growth of angular velocity. This is the reason for the appearance of the local maximum in the isolines of Fig. 6e, but also can lead to a growth of azimuthal velocity perturbations. In fact, we observe a local maximum of the azimuthal velocity perturbation amplitude on the streamlines corresponding to the flow descending along the axis (Fig. 6b). Location of the maximum near the bottom shows that the instability relates to the advection of rotation by both radial and axial components of the meridional flow, as is observed in the cases 12-15 of Table 2. A strong perturbation of the radial velocity (Fig. 6c) located at the descending part of the main vortex can be a consequence of the perturbed centrifugal force. Our description of this destabilization mechanism is supported also by the absence of strong perturbations of the axial velocity in the discussed region. We attribute this instability to the appearance of a new unstable eigenmode, which corresponds to the option (iii) described above. The latter is seen also from the stability diagram in Fig. 4a: the destabilization starts at $Re\approx175$ when a mode with $m_{cr}=4$ unstable at $Re=0$ is replaced by a mode with $m_{cr}=2$.

Comparing streamlines of the flow with and without rotation (Fig. 6a), we observe that as a result of the counter action of the buoyancy and centrifugal forces, the main convective circulation weakens. Thus, the minimal value of the stream function at $Re=0$ is –36.9, while at $Re=250$ it is –31.8. Retardation of the buoyancy convective circulation leads also to the appearance of the two counter-rotating vortices in the upper corner and in the lower part of the cylinder near its axis. A strong perturbation of the axial velocity is observed at the boundary separating the lower counter rotating vortex and the main circulation (Fig. 6d). The corresponding perturbations of two other components are weaker and are shifted downwards. It is emphasized that such a growth of perturbations at the boundary separating two counter rotating vortices was observed also in many other configurations. Nienhüser and Kuhlmann (2002) observed similar perturbations located on a similar toroidal vortex without base flow rotation. They argued that the instability is caused by a centrifugal mechanism appearing when the streamline makes a turn from the vertical descending to the radial direction.

This instability mechanism is attributed mainly to the meridional component of the flow. Our observation for the cases 16-18 shows that a slow rotation can destabilize also this mechanism. The destabilization is possibly connected with deformation of a three-dimensional perturbation of radial and axial velocities by advection around the axis caused by the non-uniform rotation. Since this mechanism results from the change of the base meridional flow, we attribute it to the option (ii). Thus, already in this case we observe a simultaneous appearance of the options (ii) and (iii).

In order to gain more understanding of the destabilization mechanisms, we observed the perturbation patterns calculated with switched off terms for the cases when it leads to stabilization (Table 2). In all the cases, the patterns remained similar to those shown in Fig. 6. We also tried to assign to the terms $\boldsymbol{e_\theta}\tilde{u}\,\partial V/\partial r$, $\boldsymbol{e_\theta}\tilde{u}\,V/r$, $\boldsymbol{e_\theta}U\,\partial(r\tilde{v})/\partial r$, and $\boldsymbol{e_\theta}W\,\partial\tilde{v}/\partial z$ an amplitude ε varying between 0 and 1, and to monitor the change of critical parameters and the perturbations when these terms are being diminished continuously. For all the above terms, we have observed continuous change of the critical Grashof number and critical frequency along with a slight, but definitely not qualitative, change in the disturbances pattern. This allows us to argue that the instability observed is generally a 3D instability of the meridional flow, for which disturbances of the azimuthal velocity component always exist. The destabilization by a slow non-uniform rotation takes place because it causes a faster growth of perturbations of the azimuthal velocity.

Analyzing the results of Table 2 for $Pr\neq0$, we notice first, that switching off the term $imV\tilde{T}$ in the linearized energy equation does not lead to stabilization in all the cases considered. We conclude that alteration of temperature perturbation by weak rotation is not a reason for the destabilization observed. We must keep in mind, however, that the perturbation pattern can change due to temperature changes in the base flow pattern.

Figure 7 shows the flow and perturbation patterns for parameters of point B in Fig. 4b. In this case the streamlines and isotherms corresponding to *Re*=0 and 200 are only slightly different (Fig. 7a and 7b). Minimal values of the stream function are –13.9 and –13.5 for *Re*=0 and *Re*=200, respectively. Since the effect of weak rotation on the temperature perturbations is already ruled out, the destabilization should be attributed to the velocity perturbations. Strong perturbations of radial and azimuthal velocities localized near the axis (Fig. 7d and 7f) can be caused by a steep increase of the azimuthal velocity observed in the same region (Fig. 7c). This assumption is supported by two additional facts. First, analysis of the leading eigenmodes for the base flow without rotation (*Re*=0) does not reveal any mode similar to the observed one, which connects this mode to the motion in the azimuthal direction, and assigns it to the option (iii). Second, switching off the term $\boldsymbol{e}_\theta \tilde{u}\, \partial V/\partial r$ in the azimuthal component of the momentum equation, rather than the Coriolis term $\boldsymbol{e}_\theta \tilde{u}\, V/r$, leads to the stabilization (Table 2, case 5). This indicates on the importance of radial non-uniformity of the azimuthal velocity for appearance of this instability. A sharp maximum of the azimuthal velocity perturbation located near the axis (Fig. 7f) also destabilizes the flow: we observe a strong destabilization when the terms $\boldsymbol{e}_\theta U\, \partial(r\tilde{v})/\partial r$ and $\boldsymbol{e}_\theta W\, \partial\tilde{v}/\partial z$ are switched off. These terms become significant because of the steep increase of the amplitude of $\tilde{v}$ near the axis.

Further, we note that switching off the term corresponding to disturbance of the centrifugal force, $2\tilde{v}V\boldsymbol{e}_r/r$, destabilizes the flow further (Table 2, case 17), so that perturbed centrifugal force acts here as a stabilizing factor. The stabilization of flow happens when we switch off the term $im\tilde{u}\, V\boldsymbol{e}_r/r$ (Table 2, cases 18 and 20) that describes advection of radial velocity disturbances in the azimuthal direction. Since perturbations of the axial velocity in the discussed region are weak, the instability should be described via an interaction between the radial and azimuthal velocity disturbances. One can imagine a feedback mechanism in which an azimuthal velocity perturbation is advected towards the axis where it grows due to conservation of the angular momentum and is advected downwards by the main convective circulation. This gives rise to the growth of radial and azimuthal velocity perturbations along the descending streamline. The maximum of the axial velocity disturbance observed in the bottom–axis corner results either from the advection of $r$- and θ-velocity perturbations downwards via continuity, or from an instability developing during the meridional streamline, turning from negative axial to positive radial direction that can be caused by the above mentioned Taylor-Couette mechanism discussed by Nienhüser and Kuhlmann (2002). In the latter case, a similar instability must be observed also without rotation. The examination of eigenmode patterns at $Re$=0 rules out the second possibility. We assume that the observed maximum of the axial velocity perturbation is also caused by the base flow rotational component. This assumption is supported also by the results of Table 2: switching off the term $imV\tilde{w}$ in the axial component of the momentum equation leads to a significant stabilization (Table 2, cases 19 and 20).

Examination of changes of the perturbation patterns when terms of the cases 5, 8, 14, 18, and 19 were reduced by introducing of an artificial amplitude ε, 0≤ε≤1, showed that the eigenmode reported in Fig. 7 disappears when the amplitude of either of the terms $\boldsymbol{e_\theta}\tilde{u}\,\partial V/\partial r$, $(U\partial(r\tilde{v})/\partial r + W\,\partial\tilde{v}/\partial z)\boldsymbol{e_\theta}$, and $(im\tilde{u}V + 2\tilde{v}V)\,\boldsymbol{e_r}/r$ is reduced below ε=0.7, 0.5, and 0.4, respectively. Switching off the terms in the cases 8 and 19, for which stabilization is also observed, does not change the leading eigenmode pattern. These observations support the above assumption of the instability mechanism: the term $\boldsymbol{e_\theta}\tilde{u}\,\partial V/\partial r$ causes the growth of the azimuthal velocity perturbation, the terms $(U\partial(r\tilde{v})/\partial r + W\,\partial\tilde{v}/\partial z)\boldsymbol{e_\theta}$ advect this perturbation along the main convective vortex, and the terms $(im\tilde{u}V + 2\tilde{v}V)\,\boldsymbol{e_r}/r$ cause the growth of perturbations of the radial velocity, which yields the necessary feedback mechanism for the appearance of sustainable oscillations.

Considering the point C on the lowest branch of the stability diagram of Fig. 4b, we observe a rather strong deformation of the purely convective flow by rotation (Fig. 8). Such a deformation is expected since we simultaneously reduce the Grashof and increase the Reynolds number. The minimum of the meridional stream function changes from –8.99 at *Re*=0 to –4.08 at *Re*=300, while another counter-rotating vortex, whose stream function maximum is 3.31, develops in the upper corner. In this case we observe the localized maximum of perturbations of axial and azimuthal velocity components located on the zero streamline separating two counter rotating vortices. The maxima of azimuthal velocity and temperature perturbations are located very close and slightly shifted into the area occupied by the weaker vortex. Two maxima of the radial velocity disturbance are shifted aside from the boundary with the larger maximum inside the weaker vortex. Here the instability should be attributed to the interaction of the two vortices. Since the splitting of the main buoyancy vortex is a clear result of superimposed convection and rotation this instability relates to option (ii). Examination of Table 2 shows, however, that excluding of several or all terms containing the base azimuthal velocity from the θ-component of the linearized momentum equation stabilizes this flow (cases 9 and 11). A switching off the term $im\tilde{u}\,V\boldsymbol{e}_r/r$ from the radial component stabilizes it even stronger (case 16). A strong stabilization is observed also when the terms $\boldsymbol{e}_\theta U\,\partial(r\tilde{v})/\partial r$ and $\boldsymbol{e}_\theta W\,\partial\tilde{v}/\partial z$ corresponding to the perturbation of azimuthal velocity are switched off (case 14). Furthermore, switching off either of the terms $im\tilde{u}\,V\boldsymbol{e}_r/r$, $\boldsymbol{e}_\theta U\,\partial(r\tilde{v})/\partial r$ and $\boldsymbol{e}_\theta W\,\partial\tilde{v}/\partial z$ or their combinations does not qualitatively change the perturbation pattern shown in Fig. 8. Thus, these terms are responsible for destabilization of the flow, but not for the appearance of the perturbation mode. Conversely, switching off all the terms containing $V$ and $\boldsymbol{e}_\theta$ (case 11) does change the disturbances pattern.

We can assume that instability of the vortices boundary is intrinsically three-dimensional, which implies a non-zero perturbation of the azimuthal velocity. Moreover, this instability sets in when a strong advection of perturbations in the azimuthal direction by the base non-uniform rotation takes place. Therefore, this case can be interpreted as a simultaneous appearance of the options (ii) and (iii).

The case of a larger Prandtl number, $Pr$=7, is illustrated in Figs. 9 and 10 corresponding to the points D and E in Fig. 4c. In this case the spoke pattern instability observed for $0 \le Re < 30$ with $m_{cr}$=10 or 11 is replaced by another one, having $m_{cr}$=1 and exhibiting a steep decrease of the critical Grashof number from $Gr_{cr} \approx 2.5\times10^4$ at $Re\approx30$ to $Gr_{cr}\approx10^4$ at $Re\approx47$. Note that this mode crosses the $Re$=0 axis at $Gr\approx4.7\times10^4$, so that in the absence of rotation it is less unstable than the spoke pattern mode. At $Re\approx47$ this mode is replaced again by the axisymmetric one ($m_{cr}$=0) that continues to even smaller values of $Gr$ reaching $Gr_{cr}\approx3000$ at $Re\approx51.5$ . With further increase of the Reynolds number the critical Grashof number slowly grows and several other mode switches take place.

Considering the example of flow and perturbation patterns shown in Fig. 9 we note that the striking feature of this case is the almost unchanged streamline and isotherm patterns corresponding to the zero and non-zero Reynolds numbers. In fact, the Reynolds number is very small, so that taking the characteristic length 10 cm and the viscosity of water $\approx10^{-6}$ $m^2/s$, $Re\approx35$ would correspond to approximately 0.2 rpm. It is really difficult to see what could change in the flow to so strongly affect its stability properties. One of possible explanations is the following. The temperature perturbation of the destabilized flow (Fig. 9c) is located near the axis in the area of unstable temperature stratification and can be driven by the Rayleigh-Bénard instability mechanism. The spoke pattern instabilities also appear due to the Rayleigh-Bénard mechanism, but their disturbances are located mainly in the thinner unstable layer closer to the cylindrical wall (Gelfgat et al., 1999). The Rayleigh-Bénard driven instabilities located near the axis were also observed in the considered configuration without rotation, but in taller cylinders (Gelfgat et al., 2000). Therefore, we observe here two competing instability modes. The examination of isotherms (Fig. 9b) shows that while unstably stratified temperature near the sidewall is unaltered by the rotation, the temperature change along the axis slightly steepens, which is seen as a slight raise of a point where two upper isotherms arrive to the axis. In Fig. 11 the axial temperature gradient of flows at several critical points are compared with those calculated at the same Grashof numbers but with zero rotation rate. We observe that the axial gradients at the instability points are always slightly steeper than those corresponding to zero rotation cases.

On the basis of the above, we can offer the following explanation of the observed destabilization. We assume that in a wide range of Grashof numbers the growth rate of the mode shown in Fig. 9 is negative but close to zero, so that this disturbance mode does not become unstable. A slow rotation of the upper boundary creates a small change in the base flow that makes the disturbance unstable. The rotation slows down the axially directed radial flow along it. Consequently, the descending flow along the axis also slows down. As a result, the convective mixing near the axis reduces, which leads to steeper axial temperature gradients. When, with the increase of the Reynolds number, the unstable temperature gradient exceeds a certain critical average value the instability sets in. We observe that at larger Grashof numbers lower axial gradient is critical (Fig. 11), which is quite expected and results from the dependence of the growth rate on the base flow.

The explanation offered assigns the observed destabilization to the option (ii), however, examination of Table 2 shows that it may be incomplete. Switching off some terms with the base rotation and their combinations (cases 2, 13, and 20), as well as a term describing the axial advection of the azimuthal velocity perturbation (case 10) can lead to a noticeable stabilization which, however, is much weaker than those observed for smaller Prandtl numbers. The pattern of the base azimuthal velocity (Fig. 9g) shows that in the upper region it is strongly advected towards the axis. The maximum of the azimuthal velocity perturbation amplitude is also located near the axis and is shifted towards the bottom by the meridional flow. As it was discussed above, such an advection tends to destabilize the flow, and we observe maxima of the meridional components perturbations located near the axis (Fig. 9d and 9e). Since the rotation is very weak, we assume that this effect is secondary, but appearance of this additional destabilization can explain why the difference between stable and unstable axial temperature gradients (Fig. 11) is so small. This assumption is supported by the observation of the disturbances pattern: in all the cases of Table 2 where stabilization was observed the patterns of leading eigenmode were similar.

The three-dimensional unsteady temperature pattern that results from the above instability mode is illustrated in Fig. 11 by a temperature isosurfaces corresponding to $T$=0.3. We observe that the isosurface forms a thin tube that rotates along the axis. Such an instability pattern was observed in experiments of Hintz et al. (2001) and Teitel et al. (2008), where it was called "cold jet" or "oscillatory jet" instability.

At the point E of Fig. 4c the critical Grashof number is already reduced to $Gr_{cr}$=5819. Since the buoyancy force in this case is much weaker, the effect or rotation becomes stronger, as is reflected in the streamline and isotherm patterns (Fig. 10a and 10b). Again, we observe a steepening of the unstable temperature gradient at the axis, which in this case leads to the so-called "cold thermals" instability, which is also of the Rayleigh-Bénard nature. Here the cold fluid is advected along the upper surface towards the axis where unstable stratification triggers the instability, appearing as a rapid descent of the cold fluid along the axis and oscillations of the main convective vortex. Note large temperature disturbances located below the upper surface and near the axis (Fig. 10c), as well as radial velocity perturbations near the upper and lower boundaries (Fig. 10d), and also the axial velocity perturbations near the axis (Fig. 10e). An additional illustration is presented in Fig. 13 by the isotherm snapshots taken from a superposition of the base flow with the perturbation. The cases shown in Fig. 13 and Fig. 5a correspond to the same branch of the neutral stability curve of Fig. 4c and, as was explained above, resemble the "cold thermals" instability observed in the experiments of Munakata & Tanasawa (1990) and of Teitel et al. (2008). The results of Table 2 do not reveal any significant dependence of the base azimuthal velocity or its perturbation on this instability. The observed instability mode becomes dominant as a result of alteration of the base meridional flow by rotation and we attribute it to the option (ii). This conclusion is supported also by results of Teitel et al. (2008) where this instability mode was observed in the Czochralski model in the absence of any base rotation both experimentally and numerically. The azimuthal velocity perturbation in this case (Fig. 10f) seems to be rather a consequence than a reason for the instability onset.

### *4.2 Destabilization versus continuous variation of the Prandtl number*

To illustrate how the described above destabilization takes place at different Prandtl numbers, we present neutral stability curves for three fixed values of the Reynolds number $Re$=0, 200 and 400, and the Prandtl number continuously varying from 0 to 10 (Fig. 14). Due to difference in the critical values and in the qualitative behavior of the curves, the graphs in Fig. 14 are divided into two frames for $Pr \leq 1$ and for $Pr \geq 1$. We observe that the destabilization at small Prandtl numbers takes place starting from a certain, not very large, value of the Reynolds number ($Re$=400), while at smaller values ($Re$=200) the critical azimuthal modes replace each other at approximately the same values of the critical Grashof number. According to arguments given in the previous Section, the destabilization at small Prandtl numbers is caused by mainly hydrodynamic mechanisms, such as interaction of counter rotating vortices and advection of angular momentum. To destabilize the base flow, these mechanisms must become strong enough, which happens at a sufficiently large value of the Reynolds number. An indication of the absence of the described destabilization effect is the appearance of the spoke pattern at non-zero *Re*. In Fig. 14a it is observed at $Re$=200 for $m_{cr}$=5 and 6.

As discussed above, at $Pr \geq 1$ the destabilization takes place mainly due to the change of the temperature distribution. This may happen at significantly smaller Reynolds numbers. In fact, for each *Pr* there exists a relatively low value of Reynolds number at which the effect is strongest, e.g. at $Re \approx 50$ in Fig. 4c. We observe also that for large Prandtl numbers the destabilization at $Re$=200 is stronger than that at $Re$=400. Note also that in the absence of rotation the instability results in a spoke pattern. The critical Grashof numbers at $Re$=0 are very close for $10 \leq m \leq 15$ (Fig. 14b). As discussed above, at a relatively weak rotation the modes related to oscillating jet or cold plumes instability become most unstable and replace the spoke pattern mode, thus leading to destabilization. The oscillating jet and cold plumes modes are characterized by a smaller azimuthal wavenumber, which is seen on the curves corresponding to $Re$=200 and 400 in Fig. 14b.

### *4.3 Destabilization of convection under rotating magnetic field effect*

Consider now destabilization of a convective flow by the rotating magnetic field (Fig. 2c). Without the RMF effect the flow becomes unstable at $Gr_{cr} \approx 10^5$. The corresponding flow and perturbation patterns are shown in Fig. 15. At $Ta \approx 1600$, which corresponds to a rather weak RMF effect (Grants & Gerbeth, 2002), the neutral curve branch started at $Ta$=0 is replaced by another one steeply descending and reaching the minimal critical value of the Grashof number approximately $1.5\times10^4$ at $Ta\approx1300$ (Fig. 2c). Along this neutral branch, the instability sets in due to increase of *Ta* rather than due to a change of *Gr*. Flow and perturbation patterns corresponding to this instability mode are shown in Fig. 16. The first observation is that the meridional flow is almost unaltered by RMF (Fig. 16a and 16b), which indicates again on the RMF weakness. Contrarily, isolines of the azimuthal velocity are significantly different when affected or not affected by the convective circulation (Fig. 16c). Gradually increasing the Grashof number from $Gr=0$ to $2\times10^4$ we observe that the maximum of *V* first shifts upwards, then toward the axis, and then descends downwards, so that the isolines become almost symmetric, with respect to the $z=1$ plane, at $Gr=2\times10^4$. Since the non-zero azimuthal velocity is the only significant difference from the $Ta=0$ case, we assume that this is also the main reason for destabilization. In fact, by switching of one or several of the terms $im\frac{\tilde{u}V}{r}\boldsymbol{e}_r$, $2\frac{\tilde{v}V}{r}\boldsymbol{e}_r$, and $imV\tilde{w}\boldsymbol{e}_z$, belonging to the meridional part of the linearized momentum equation, we observe a considerable stabilization of the flow. At the same time switching of the *V*-depending terms in the azimuthal component leads only to minor changes of the critical parameters.  The examination of eigenmodes at $Ta=0$ shows that there is no disturbance similar to that shown in Fig. 16d-16g when the RMF is absent. This makes this instability belonging to the option (iii).

The observed instability cannot be related to the Taylor-Couette mechanism because the meridional velocity perturbations are large closer to the axis (Fig. 16d and 16c), where azimuthal velocity decreases with the decrease of the radius, thus making the local rotation in this region stable. We conclude that the instability sets in due to advection of base rotation by meridional velocity disturbances, and propose the following description. An appearance of a negative disturbance of the radial velocity at a point where its amplitude attains the maximum can bring a particle with a larger rotational moment towards the axis, where the rotation will increase. This will increase the centrifugal force and, consequently, a positive perturbation of the radial velocity. The latter will happen during advection of the particle downwards by the main vortex, and will happen already at another azimuthal angle. An increased centrifugal force will increase the positive radial velocity in the lower part of the cylinder, which speeds up the main convective vortex. The perturbed vortex will create a negative perturbation of the radial velocity in the upper part of the flow and, therefore, there will be the feedback mechanism leading to the meridional flow oscillations.

### *4.4 Destabilization by rotation in Czochralski model flow*

Now we return to the Czochralski model flow that motivated this study. The full formulation of the problem can be found in Crnogorac et al. (2008) and Gelfgat (2008). The crucible radius is chosen as the characteristic length, and therefore definitions of the Grashof and Reynolds number remain unchanged. The flow is described by the Boussinesq equations (1)-(3) with no-slip boundary conditions at the bottom and the sidewall of the crucible, while the temperature there is prescribed according to the experiment of Schwabe et al. (2004)

$$\text{at } r = 1: \qquad v_r = v_\theta = v_z = 0, \quad T = 1 \tag{11}$$

$$\text{at } z = 0: \qquad v_r = v_z = v_\theta = 0, \quad T = 0.8571 + 0.1429r^2 \tag{12}$$

The central part of the upper surface touches the rotating cold crystal dummy kept at the lower temperature. The remaining part of the upper surface is cooled by a convective flow of air above it and is subject to the thermocapillary force. This reads

$$\text{at } z = A \text{ and } r \le \frac{R_{crystal}}{Rcrucible}: \qquad v_r = v_z = 0,\ v_\theta = Re_{crystal} r \tag{13}$$

$$\text{at } z = A \text{ and } r > \frac{R_{crystal}}{Rcrucible}: \quad v_z = 0, \quad \frac{\partial v_r}{\partial z} = -MaPr\frac{\partial T}{\partial r}, \quad \frac{\partial v_\theta}{\partial z} = -MaPr\frac{\partial T}{r\partial \theta} \tag{14}$$

The flow is driven by buoyancy, thermocapillarity and rotation, which are characterized by the Grashof, Marangoni and Reynolds numbers. The Marangoni number is defined as $Ma = \gamma\Delta TR/\nu\alpha$, where $\gamma = |d\sigma/dt|$ is assumed to be a constant and σ is the surface tension coefficient. The working liquid is $NaNO_3$ with *Pr*=9.2. Since both the Grashof and Marangoni numbers depend on the characteristic temperature difference we calculate $Ma$=586 $\Delta T/Pr$ and $Gr=1.9\times10^5\Delta T$ and use $\Delta T$ as a critical parameter.

Figure 17 shows changes of the flow and perturbation patterns along the neutral stability curve of Fig. 1a. Since the instability in this case is axisymmetric ($m_{cr}$=0), we plot the stream function and its perturbation instead of the two meridional velocity components. As above, in the case of our characteristic problem at *Pr*=7, we observe that with the increase of rotation the main convective circulation weakens, which leads to a steepening of the axial unstable temperature gradient near the axis. This leads to strong temperature perturbations that develop below the cold crystals and descend with the flow along the axis. The snapshots of isotherms shown in Figs. 18 and 19 for the points $a$ and $c$ of Fig. 1a, respectively, show similar oscillations of cold thermals that descend along the axis. Examination of the case of Teitel et al. (2008) shown in Fig. 1b shows similar perturbation patterns and similar time-dependence. We conclude that the destabilization observed for the large Prandtl number Czochralski melt flow has the same nature as the one observed for the simplified characteristic problem. Rotation of the crystal causes a retardation of the main convective circulation, which leads to a formation of an unstably stratified layer beyond the crystal. This layer is destabilized by the Rayleigh-Bénard mechanism, which is mainly defined by the layer thickness. With the increase of the rotation rate (the Reynolds number) formation of the unstable layer takes place at a lower Grashof number, thus resulting in the destabilization of this convective flow by rotation.

## 4 Conclusions

We have studied the destabilization of buoyancy convection flows by a weak rotation, which was observed in the Czochralski model flows experimentally and numerically and motivated this study. Seeking for a simplified problem that reveals similar destabilization we showed that thermal boundary conditions leading to the ascending of hot fluid along the lateral boundary and descending of the cold fluid along the axis together with the rotating upper boundary, represent the needed characteristic problem. Therefore, a combination of previously studied problems of convection in a cylinder with parabolic sidewall temperature profile and swirling flow in a cylinder with a rotating lid was taken as a characteristic problem for further consideration. Several characteristic cases of destabilization were studied by observation of their flow and leading disturbance patterns, and by switching off some of the terms of the linearized stability equations (Table 2). The main conclusions are summarized in Table 3. To show that the destabilization of convection by a weak non-uniform rotation can extend also to other flow configurations we supplied an example of convective flow interacting with RMF driven rotation.

Based on all above observations we can expect the destabilization effect when buoyancy and centrifugal forces tend to create meridional vortices of opposite direction. This leads to two rather obvious effects. The first one is a splitting of main convective circulation into several vortices with an unstable boundary between them that causes instability. The second effect is retardation of convective mixing and creation of unstably stratified regions where Rayleigh-Bénard instability mechanism sets in. The third effect observed is connected with the advection of the angular momentum and its perturbation towards the axis, which leads to growth of the azimuthal velocity with a consequent local growth of the centrifugal force. The latter increases the base radial velocity and its perturbation, thus intensifying advection and creating a positive feedback. This effect is found to be dominant, e.g., in the case of RMF destabilization of convection. In some other cases it was observed as an additional destabilization that enhances the two previous effects.

It should be mentioned also that starting the discussion on the mechanisms that lead to the convective flow destabilization by rotation we defined three main cases, which are (i) alteration of a leading disturbance by rotation making it unstable, (ii) alteration of the base meridional flow pattern that leads to a destabilization of one of the leading eigenmodes, and (iii) an appearance of a new eigenmode caused by the base rotational motion. Surprisingly, we observed cases (ii) and (iii) existing separately, as well as interacting. However, we did not observe case (i).

**Acknowledgement**

This study was supported by the German-Israeli Foundation, grant No. I-954 -34.10/2007.

Table 1. Temperature boundary conditions, values of the governing parameters, and maximal and minimal stream function values for the frames and inserts of Figs. 2-4.

| Figure | $z=0$ | $z=1$ | $r=1$ | $A$ | $Pr$ | $Gr$ | $Re_{top}$ | $Re_{bottom}$ | $Ta$ | $\psi_{min}$ | $\psi_{max}$ |
|---|---|---|---|---|---|---|---|---|---|---|---|
| 2a | $T=1-r^2$ | $T=0$ | $\frac{\partial T}{\partial r}=0$ | 1 | 7 | $10^5$ | 0 | 0 | 0 | 0 | 2.536 |
| 2a | $T=1-r^2$ | $T=0$ | $\frac{\partial T}{\partial r}=0$ | 1 | 7 | $2.7\cdot10^4$ | 200 | 0 | 0 | 0 | 3.280 |
| 2b | $T=\begin{cases}1, r\le 0.5\\ 0, r>0.5\end{cases}$ | $T=0$ | $\frac{\partial T}{\partial r}=0$ | 1 | 0.7 | $3.4\cdot10^5$ | 0 | 0 | 0 | 0 | 13.683 |
| 2b | $T=\begin{cases}1, r\le 0.5\\ 0, r>0.5\end{cases}$ | $T=0$ | $\frac{\partial T}{\partial r}=0$ | 1 | 0.7 | $1.08\cdot10^5$ | 200 | 0 | 0 | 0 | 9.031 |
| 2c | $T=0$ | $T=0$ | $T=4z(1-z)$ | 1 | 0.015 | $10^5$ | 0 | 0 | 0 | -26.885 | 0 |
| 2c | $T=0$ | $T=0$ | $T=4z(1-z)$ | 2 | 0.015 | $2\cdot10^4$ | 0 | 0 | 1020 | -7.985 | 0 |
| 2c | $T=0$ | $T=0$ | $T=4z(1-z)$ | 2 | 0.015 | $10^5$ | 0 | 0 | $10^4$ | -15.946 | 0.938 |
| 3a | $T=0$ | $T=0$ | $T=4z(1-z)$ | 1 | 0.015 | $2.57\cdot10^5$ | 0 | 0 | 0 | -72.156 | 0.0413 |
| 3a | $T=0$ | $T=0$ | $T=4z(1-z)$ | 1 | 0.015 | $3.19\cdot10^5$ | 0 | 300 | 0 | -68.099 | 0.00552 |
| 3a | $T=0$ | $T=0$ | $T=4z(1-z)$ | 1 | 0.015 | $3.6\cdot10^5$ | 0 | 600 | 0 | -62.237 | 0.00194 |
| 3b | $T=0$ | $T=0$ | $T=4z(1-z)$ | 1 | 0.7 | $3.9\cdot10^5$ | 0 | 0 | 0 | -14.499 | 0.297 |
| 3b | $T=0$ | $T=0$ | $T=4z(1-z)$ | 1 | 0.7 | $6\cdot10^5$ | 0 | 175 | 0 | -16.690 | 0.700 |
| 3b | $T=0$ | $T=0$ | $T=4z(1-z)$ | 1 | 0.7 | $6\cdot10^5$ | 0 | 400 | 0 | -16.318 | 0.725 |

Table 1. (continued)

| Figure | $z = 0$ | $z = 1$ | $r = 1$ | $A$ | $Pr$ | $Gr$ | $Re_{top}$ | $Re_{bottom}$ | $Ta$ | $\psi_{min}$ | $\psi_{max}$ |
|---|---|---|---|---|---|---|---|---|---|---|---|
| 3c | $T = 0$ | $T = 0$ | $T = 4z(1 - z)$ | 1 | 7 | $3.55 \cdot 10^4$ | 0 | 0 | 0 | -1.532 | 0.0355 |
| 3c | $T = 0$ | $T = 0$ | $T = 4z(1 - z)$ | 1 | 7 | $4.4 \cdot 10^4$ | 0 | 200 | 0 | -1.650 | 0.0414 |
| 4a | $T = 0$ | $T = 0$ | $T = 4z(1 - z)$ | 1 | 0.015 | $3.1 \cdot 10^5$ | 150 | 0 | 0 | -65.118 | 0.0243 |
| 4a | $T = 0$ | $T = 0$ | $T = 4z(1 - z)$ | 1 | 0.015 | $1.1 \cdot 10^5$ | 300 | 0 | 0 | -26.920 | 0.702 |
| 4a | $T = 0$ | $T = 0$ | $T = 4z(1 - z)$ | 1 | 0.015 | $4 \cdot 10^4$ | 600 | 0 | 0 | -5.976 | 1.656 |
| 4b | $T = 0$ | $T = 0$ | $T = 4z(1 - z)$ | 1 | 0.7 | $7 \cdot 10^5$ | 100 | 0 | 0 | -17.545 | 1.003 |
| 4b | $T = 0$ | $T = 0$ | $T = 4z(1 - z)$ | 1 | 0.7 | $5 \cdot 10^5$ | 146 | 0 | 0 | -15.636 | 0.621 |
| 4b | $T = 0$ | $T = 0$ | $T = 4z(1 - z)$ | 1 | 0.7 | $3.57 \cdot 10^5$ | 260 | 0 | 0 | -12.971 | 0.786 |
| 4b | $T = 0$ | $T = 0$ | $T = 4z(1 - z)$ | 1 | 0.7 | $8.8 \cdot 10^4$ | 400 | 0 | 0 | -2.836 | 5.504 |
| 4c | $T = 0$ | $T = 0$ | $T = 4z(1 - z)$ | 1 | 7 | $3.8 \cdot 10^4$ | 24 | 0 | 0 | -1.454 | 0.0617 |
| 4c | $T = 0$ | $T = 0$ | $T = 4z(1 - z)$ | 1 | 7 | $10^4$ | 47 | 0 | 0 | -0.807 | 0.0172 |
| 4c | $T = 0$ | $T = 0$ | $T = 4z(1 - z)$ | 1 | 7 | $4.3 \cdot 10^3$ | 70 | 0 | 0 | -0.292 | 0.365 |
| 4c | $T = 0$ | $T = 0$ | $T = 4z(1 - z)$ | 1 | 7 | $1.28 \cdot 10^4$ | 200 | 0 | 0 | -0.288 | 2.413 |

Table 2. Effect of azimuthal component of the base flow on the critical Grashof number. (+) – term is included, (–) – term is excluded.

| case | $\tilde{u}\frac{\partial V}{\partial r}\boldsymbol{e}_\theta$ | $\frac{\tilde{u}V}{r}\boldsymbol{e}_\theta$ | $im\tilde{v}V\boldsymbol{e}_\theta$ | $\tilde{w}\frac{\partial V}{\partial z}\boldsymbol{e}_\theta$ | $\frac{U}{r}\frac{\partial(r\tilde{v})}{\partial r}\boldsymbol{e}_\theta$ | $W\frac{\partial \tilde{v}}{\partial z}\boldsymbol{e}_\theta$ | $im\frac{\tilde{u}V}{r}\boldsymbol{e}_r$ | $2\frac{\tilde{v}V}{r}\boldsymbol{e}_r$ | $imV\tilde{w}\boldsymbol{e}_z$ | $imV\tilde{T}$ | $Pr$=0, $m$=2, $Re$=250 | $Pr$=0.7, $m$=1, $Re$=200 | $Pr$=0.7, $m$=2, $Re$=300 | $Pr$=7, $m$=1, $Re$=34.8 | $Pr$=7, $m$=0, $Re$=50 |
|---|---|---|---|---|---|---|---|---|---|---|---|---|---|---|---|
| 1 | + | + | + | + | + | + | + | + | + | + | 119000 | 349700 | 86220 | 20000 | 5819 |
| 2 | + | + | + | – | + | + | + | + | + | + | 158900 | 301200 | 90210 | 34500 | 6704 |
| 3 | + | + | – | + | + | + | + | + | + | + | 147800 | 204200 | 112800 | 13600 | 5819 |
| 4 | + | + | – | – | + | + | + | + | + | + | 175300 | 192800 | 103100 | 18360 | 6704 |
| 5 | – | + | + | + | + | + | + | + | + | + | 228000 | 764200 | 100400 | 21260 | 4187 |
| 6 | + | – | + | + | + | + | + | + | + | + | 50600 | 52310 | 64930 | 20950 | 5187 |
| 7 | – | – | + | + | + | + | + | + | + | + | 247000 | 54350 | 68320 | 23000 | 2559 |
| 8 | + | + | + | + | – | + | + | + | + | + | 98500 | 480000 | 75400 | 18800 | 5962 |
| 9 | – | – | – | + | – | + | + | + | + | + | 387000 | 24000 | 643000 | 13500 | 3031 |
| 10 | + | + | + | + | + | – | + | + | + | + | 136000 | 996000 | 653000 | 27100 | 5883 |
| 11 | – | – | – | – | + | + | + | + | + | + | 243300 | 243300 | 285493 | 18850 | 2672 |
| 12 | – | – | – | – | – | + | + | + | + | + | 399000 | 29000 | 348000 | 18600 | 4113 |
| 13 | – | – | – | – | + | – | + | + | + | + | 213000 | 39000 | 348000 | 35500 | 4112 |
| 14 | + | + | + | + | – | – | + | + | + | + | 268000 | 853000 | 665000 | 23800 | 4113 |
| 15 | – | – | – | – | – | – | + | + | + | + | 419000 | 30500 | 252000 | 26000 | 4112 |
| 16 | + | + | + | + | + | + | – | + | + | + | 224200 | 719900 | 507900 | 28000 | 5819 |
| 17 | + | + | + | + | + | + | + | – | + | + | 255000 | 82440 | 72940 | 25070 | 4113 |
| 18 | + | + | + | + | + | + | – | – | + | + | 197500 | 849500 | 90220 | 23260 | 4113 |
| 19 | + | + | + | + | + | + | + | + | – | + | 106100 | 940100 | 65370 | 19130 | 5819 |
| 20 | + | + | + | + | + | + | – | – | – | + | 119000 | 966200 | 48550 | 29080 | 4113 |
| 21 | + | + | + | + | + | – | + | + | + | – | 119000 | 298500 | 30700 | 15480 | 5819 |
| 22 | – | – | – | – | + | + | – | – | – | – | 125900 | 803200 | 61150 | 34420 | 4113 |
| 23 | – | – | – | – | – | – | – | – | – | – | 363000 | 950000 | 726000 | 32700 | 4113 |

Table 3. Different cases of destabilization of non-isothermal flow in a vertical cylinder with parabolic temperature profile at the side wall and rotating top discussed in the Section 4.1 .

| Parameters | Location on stability diagram | Case of destabilization | Instability sources |
|---|---|---|---|
| $Pr$=0, $Re$=250, $Gr$=1.19×10$^{5}$ | point A, Fig. 4a | (ii) + (iii) | advection of the base angular momentum and its perturbation towards the axis; interaction of counter-rotating vortices |
| $Pr$=0.7, $Re$=200, $Gr$=3.497×10$^{5}$ | point B, Fig. 4b | (iii) | advection of the base angular momentum and its perturbation towards the axis |
| $Pr$=0.7, $Re$=300, $Gr$=8.622×10$^{4}$ | point C, Fig. 4b | (ii) + (iii) | interaction of counter-rotating vortices |
| $Pr$=7, $Re$=34.8, $Gr$=2.0×10$^{4}$ | point D, Fig. 4c | (ii) | unstable stratification along the axis leading to "oscillating jet" instability |
| $Pr$=7, $Re$=50, $Gr$=5819 | point E, Fig. 4c | (ii) | unstable stratification along the axis leading to "cold thermals" instability |

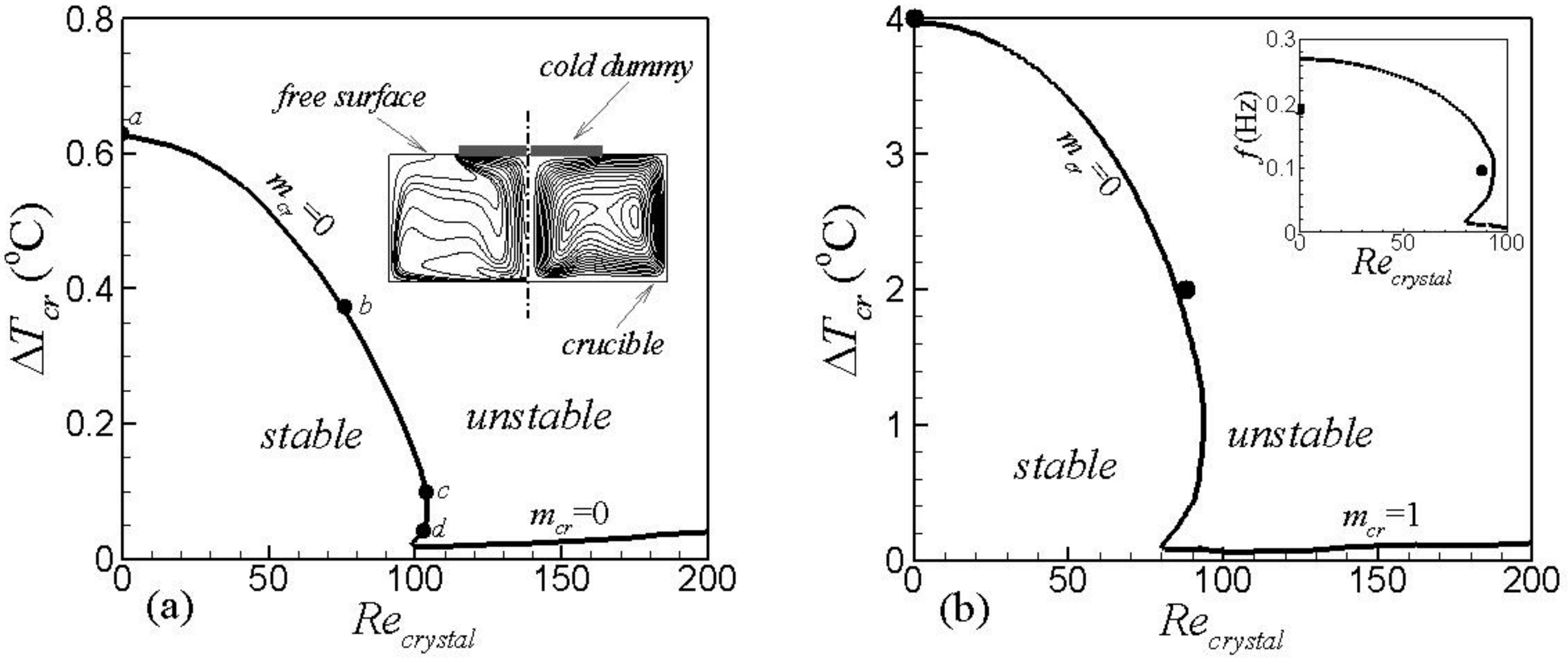

Figure 1. Neutral stability curves for two models of Czochralski melt flow driven by buoyancy, thermocapillarity and rotation. (a) – configuration of the experiment of Schwabe *et al*. (2004); the insert shows the streamlines (right frame) and isotherms (left frame) at $\Delta T = 0.5$. (b) configuration of the experiment of Teitel *et al.* (2008); symbols correspond to experimentally measured critical points, the insert shows frequency of flow oscillations at the critical points (curve) and experimentally measured frequencies.

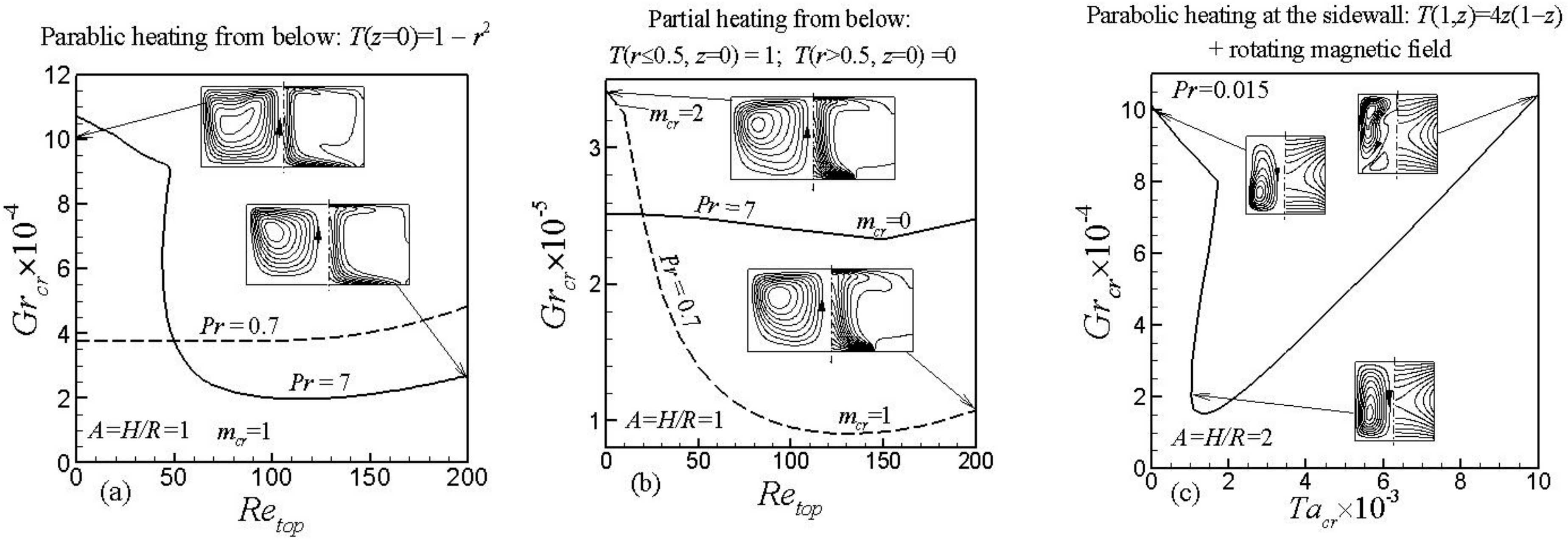

Figure 2. Neutral stability curves for different flows in a vertical cylinder driven by buoyancy convection and rotation. Inserts show streamlines (left frames) and isotherms (right frames) in several characteristic points. (a) – cylinder with a parabolic temperature profile at the bottom and rotating top, (b) cylinder with partially heated bottom and rotating top, (c) cylinder with a parabolic temperature profile at the sidewall under effect of rotating magnetic field. The isotherms are equally spaced between the values 0 and 1. Streamlines are equally spaced between the minimal and maximal values reported in Table 1.

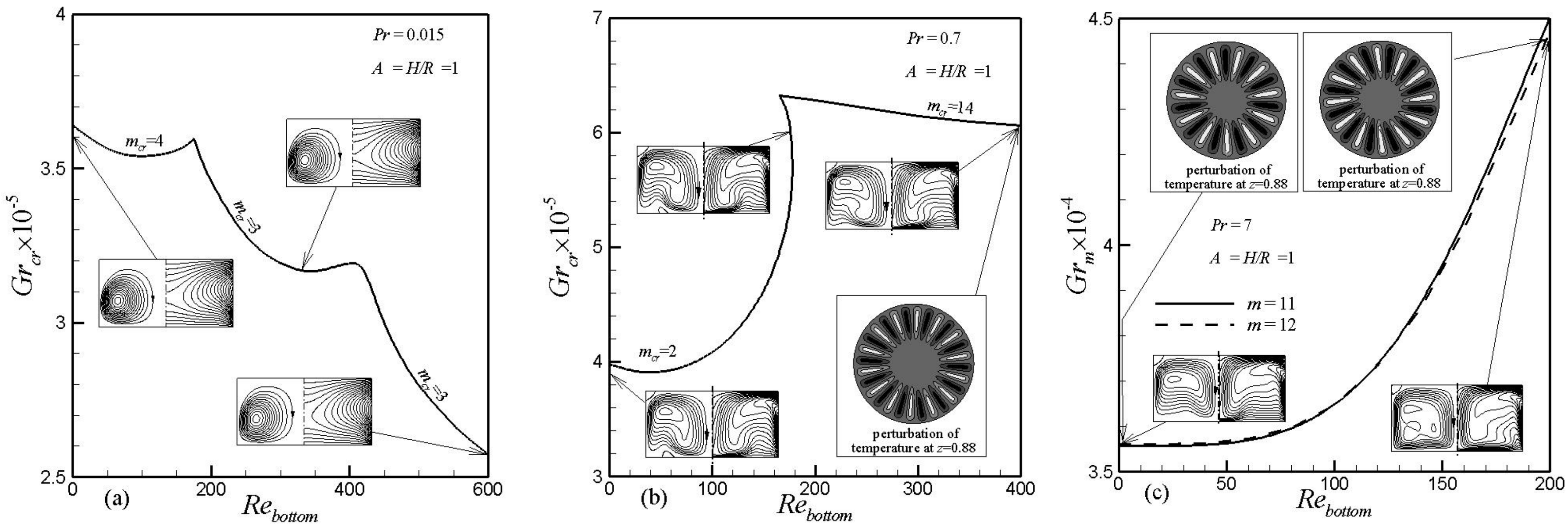

Figure 3. Neutral stability curves for flow in a vertical cylinder with parabolic temperature profile at the sidewall and rotating bottom. Inserts show streamlines (left frames) and isotherms (right frames) in several characteristic points. The isotherms are equally spaced between the values 0 and 1. Streamlines are equally spaced between the minimal and maximal values reported in Table 1. Zero streamline is added where it exists. Additional inserts in frames (b) and (c) show temperature perturbation for the spoke pattern instabilities.

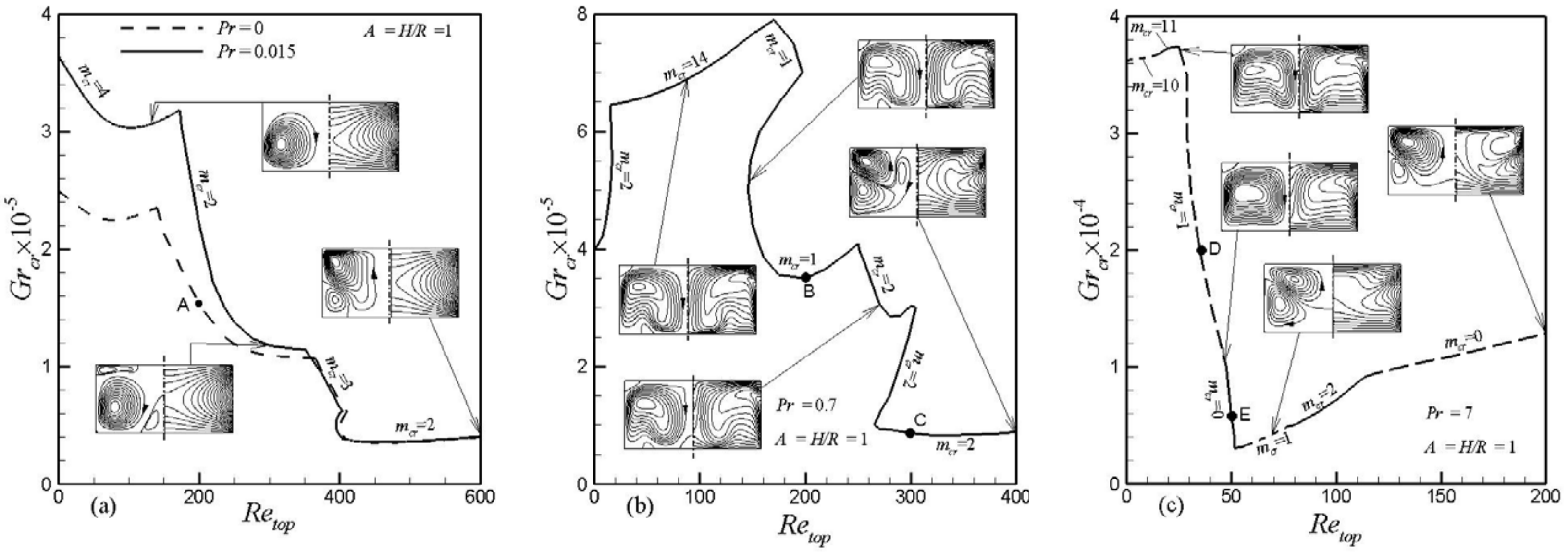

Figure 4. Neutral stability curves for flow in a vertical cylinder with parabolic temperature profile at the sidewall and rotating top. Inserts show streamlines (left frames) and isotherms (right frames) in several characteristic points. The isotherms are equally spaced between the values 0 and 1. Streamlines are equally spaced between the minimal and maximal values reported in Table 1. Zero streamline is added where it exists. The neutral curve of the frame (c) is shown by solid and dashed lines to distinguish between the parts corresponding to different critical azimuthal wavenumbers $m_{cr}$.

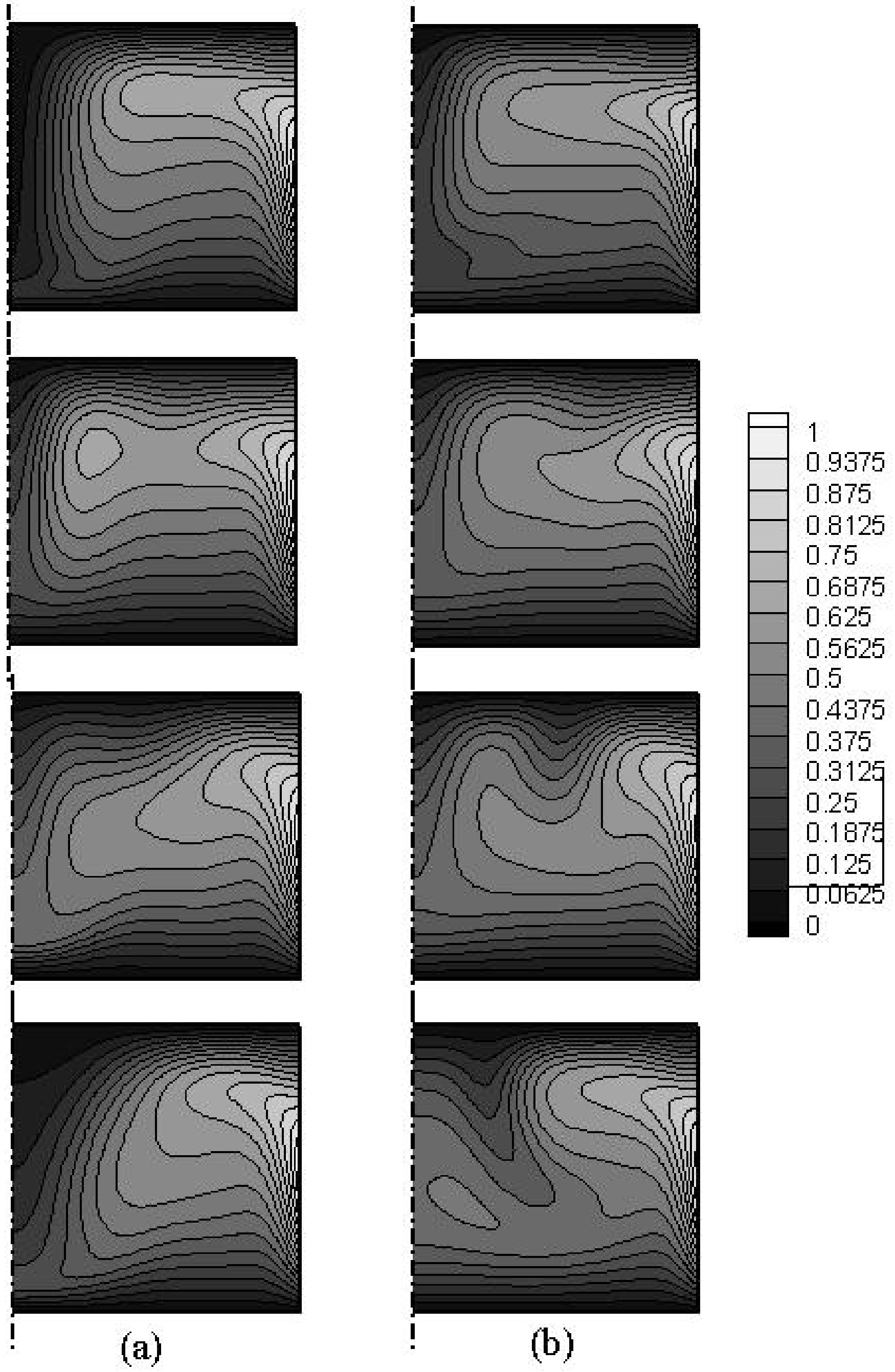

Fig. 5. Four equally distanced snapshots of supercritical oscillatory state during one period of oscillations. Flow with parabolic temperature profile at the sidewall with rotating top at $Pr$=7 (Fig. 4c). (a) Superposition of the base flow with the most unstable perturbation, $Gr_{cr}$=$10^4$, $Re_{cr}$=47.51, $m_{cr}$=0. (b) Fully nonlinear calculation for $Gr$=$10^4$, $Re$=60, $m$=0.

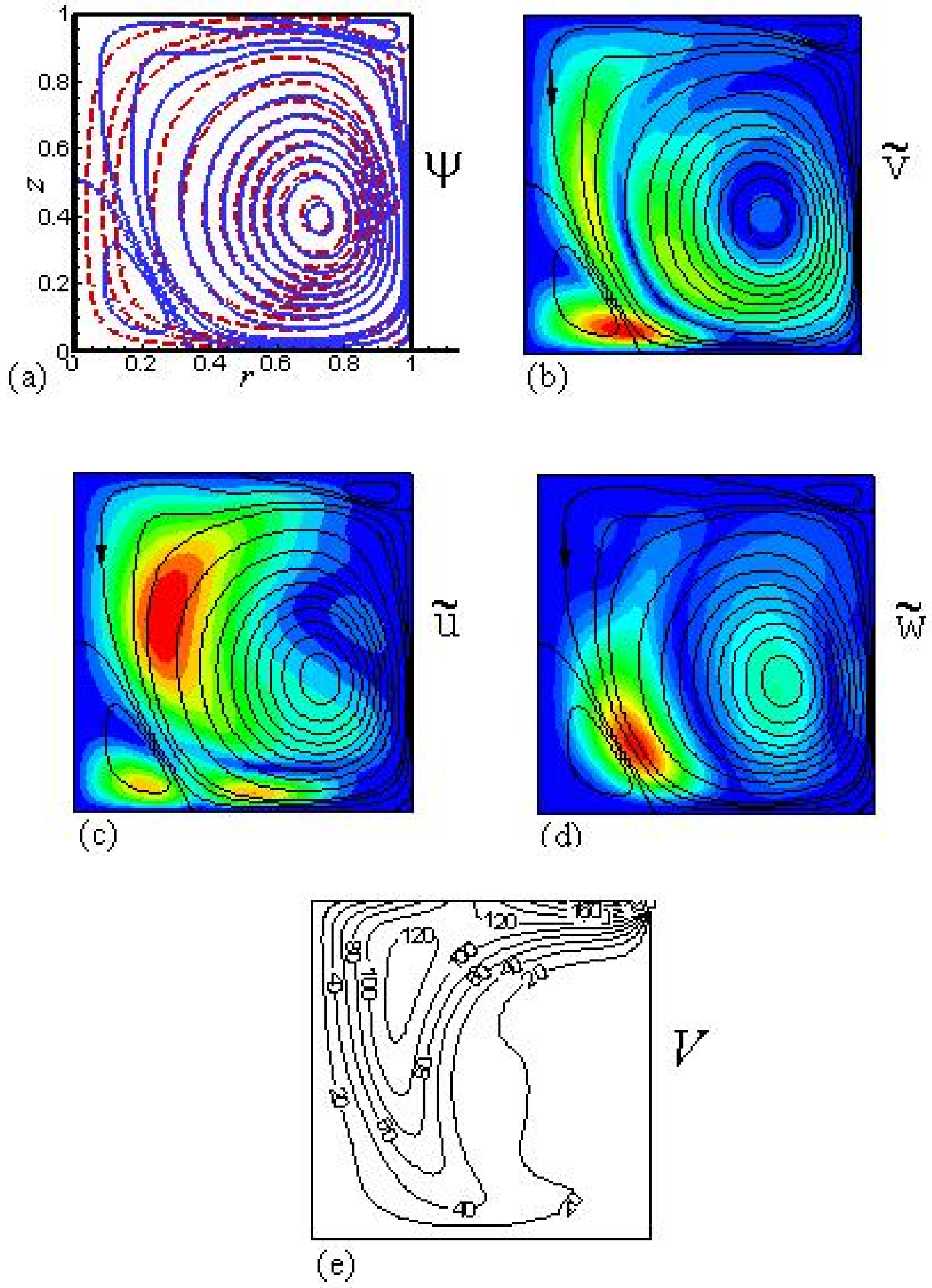

Fig. 6. Flow and perturbation patterns for point A in Fig. 4a. $Pr$=0, $Re$=250, $Gr_{cr}=1.19\times10^5$. Instability for the azimuthal wavenumber $m$=2. (a) streamlines of the considered flow (blue solid line) and of the flow at $Re$=0 at the same $Gr$ (red dash line). Streamlines are evenly distributed between the values of 0 and -36 (global minimum); levels ±0.1 and -0.5 are added additionally. Frames (b), (c), and (d) show absolute value of the most unstable perturbation of the azimuthal, radial and axial velocities, respectively. Frame (e) shows isolines of the azimuthal component of the base flow.

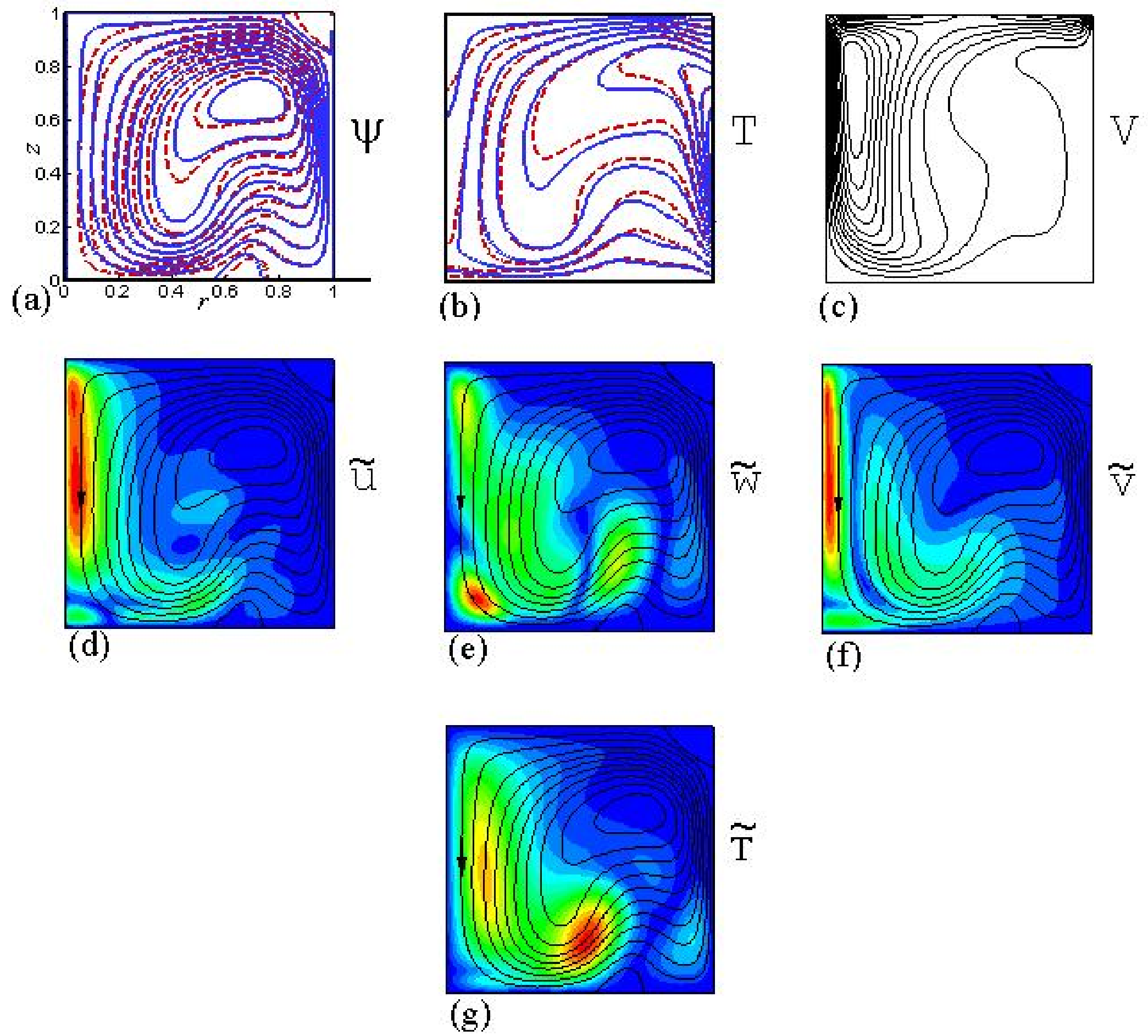

Fig. 7. Flow and perturbation patterns for point B in Fig. 4b. $Pr$=0.7, $Re$=200, $Gr_{cr}=3.497\times10^5$. Instability for the azimuthal wavenumber $m$=1. (a) streamlines, (b) isotherms of the considered flow (blue solid line) and of the flow at $Re$=0 at the same $Gr$ (red dash line). Streamlines are evenly distributed between the values of 0 and -14 (global minimum); level -0.5 is added additionally. Isotherms are evenly distributed between the values 0 and 1. (c) - isolines of azimuthal velocity evenly distributed between 0 and 240. Frames (d), (e), (f) and (g) show absolute value of the most unstable perturbation of radial, axial and azimuthal velocities, and the temperature, respectively.

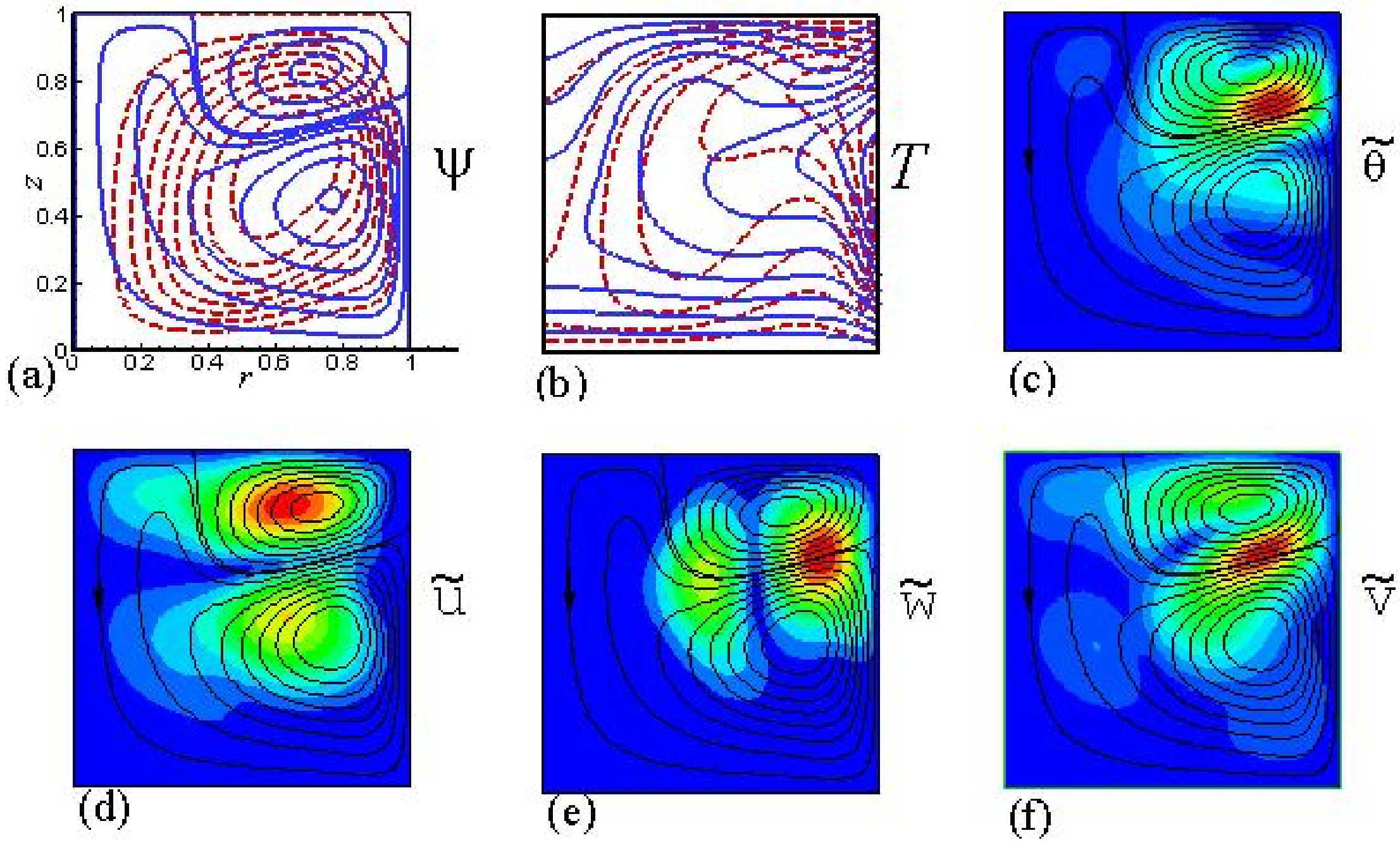

Fig. 8. Flow and perturbation patterns for point C in Fig. 4b. $Pr$=0.7, $Re$=300, $Gr_{cr}$=8.622×$10^4$. Instability for the azimuthal wavenumber $m$=2. (a) streamlines and (b) isotherms of the considered flow (blue solid line) and of the flow at $Re$=0 at the same $Gr$ (red dash line). Streamlines are evenly distributed between the values of 3 and -8 (global minimum). Isotherms are evenly distributed between the values 0 and 1. Frames (c), (d), (e) and (f) show absolute value of the most unstable perturbation of the temperature and radial, axial and azimuthal velocities, respectively.

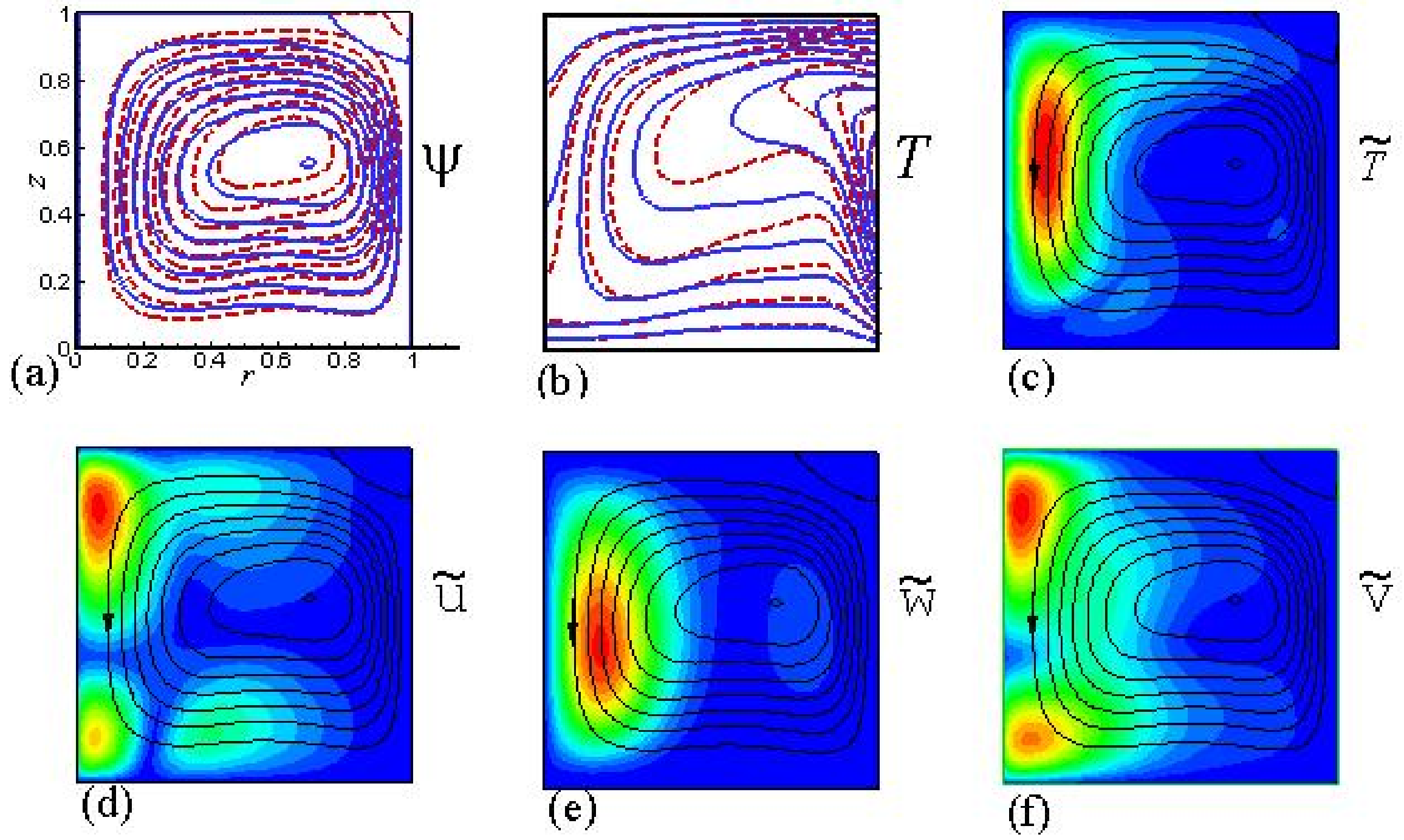

Fig. 9. Flow and perturbation patterns for point D in Fig. 4c. $Pr$=7, $Re$=34.8, $Gr_{cr}$=2.0×$10^4$. Instability for the azimuthal wavenumber $m$=1. (a) streamlines and (b) isotherms of the considered flow (blue solid line) and of the flow at $Re$=0 at the same $Gr$ (red dash line). Streamlines are evenly distributed between the values of 0 and -1.3 (global minimum). Isotherms are evenly distributed between the values 0 and 1. Frames (c), (d), (e) and (f) show absolute value of the most unstable perturbation of the temperature and radial, axial and azimuthal velocities, respectively. Frame (g) shows isolines of the azimuthal component of the base flow.

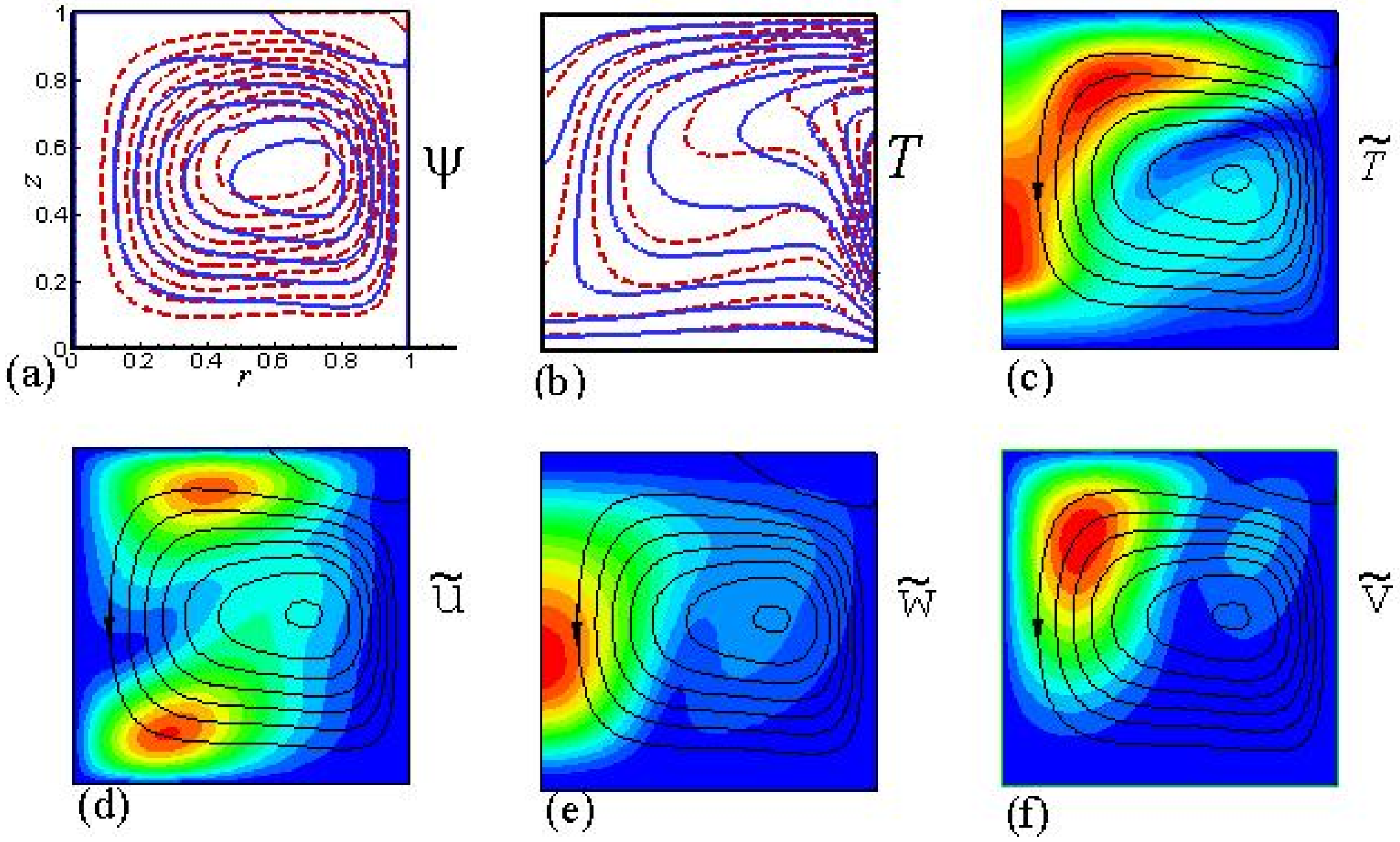

Fig. 10. Flow and perturbation patterns for point E in Fig. 4c. *Pr*=7, *Re*=50, $Gr_{cr}$=5819. Instability for the azimuthal wavenumber *m*=0. (a) streamlines and (b) isotherms of the considered flow (blue solid line) and of the flow at *Re*=0 at the same *Gr* (red dash line). Streamlines are evenly distributed between the values of 0 and -0.8 (global minimum), and a level -0.02 is added additionally. Isotherms are evenly distributed between the values 0 and 1. Frames (c), (d), (e) and (f) show absolute value of the most unstable perturbation of the temperature and radial, axial and azimuthal velocities, respectively.

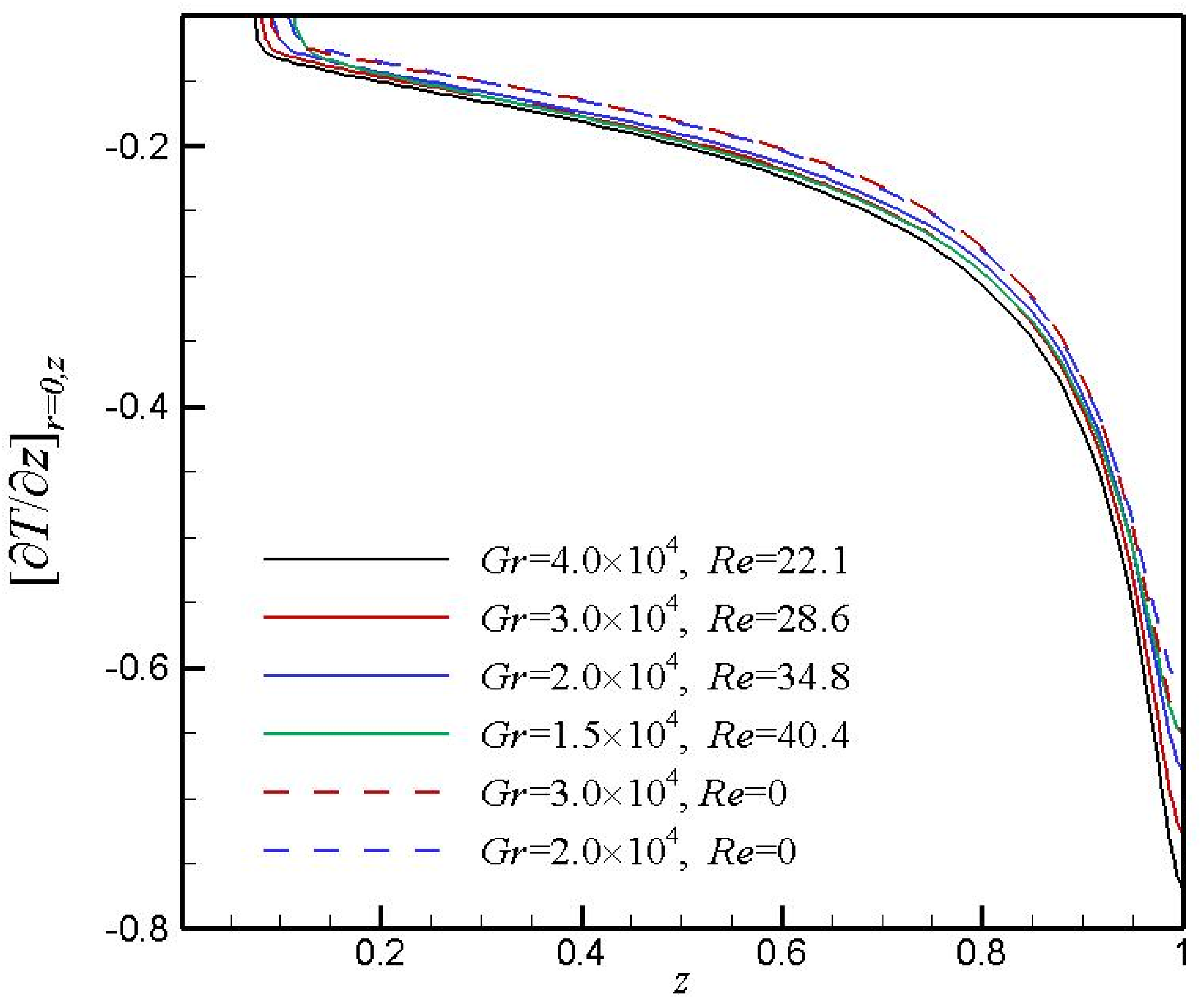

Fig. 11. Axial temperature gradients for several flows at the critical points corresponding to the descending $m$=1 branch of Fig. 4c, and several flows at the zero rotation.

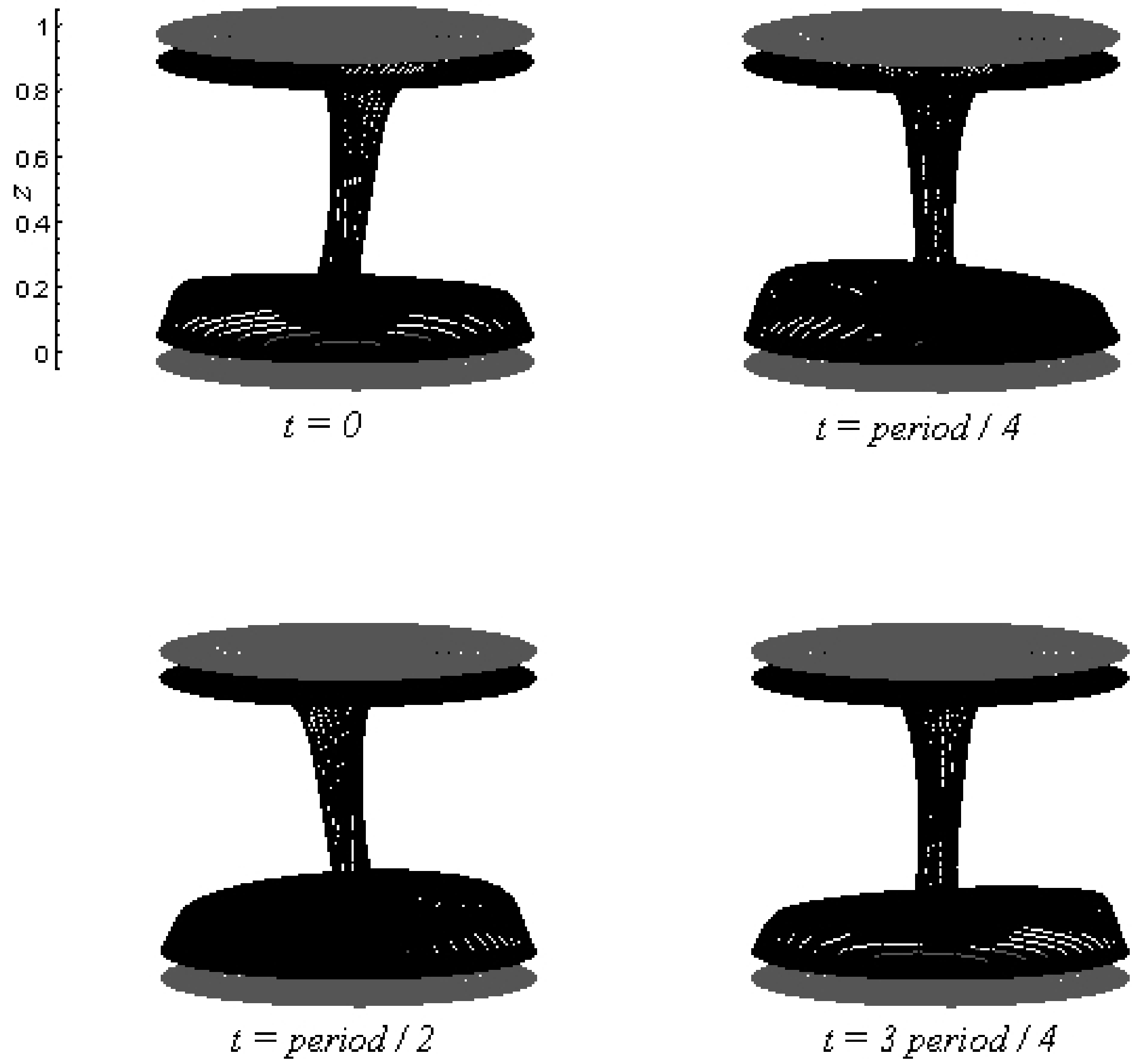

Fig. 12. Oscillations of the temperature isosurface $T$=0.3 corresponding to the instability at $Pr$=7, $Re$=34.8, $Gr=2\times10^4$.

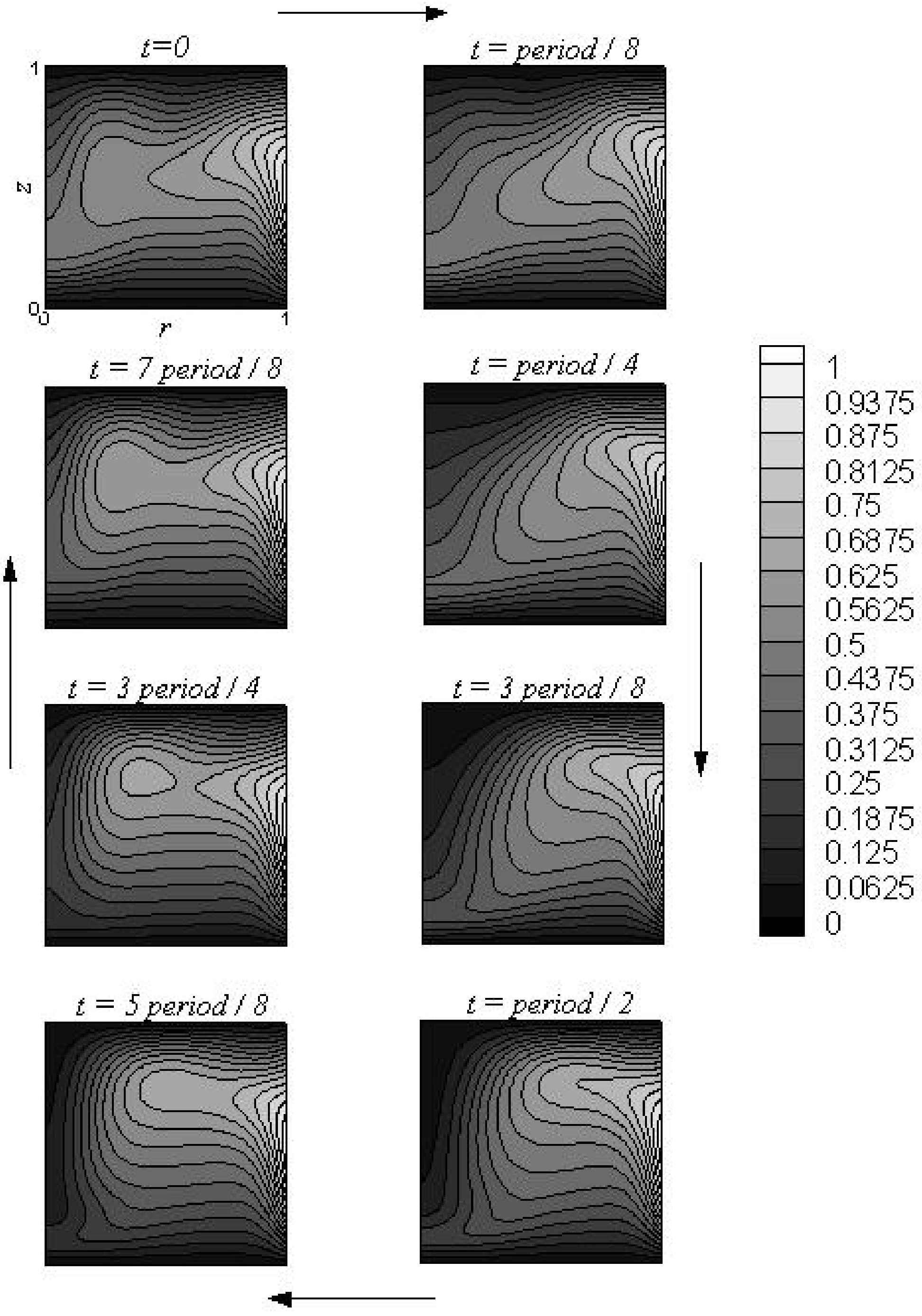

Fig. 13. Eight equally distanced snapshots of isotherms of supercritical oscillatory state during one period of oscillations. Flow with parabolic temperature profile at the sidewall with rotating top (Fig. 4c, point E) at $Pr$=7, $Gr_{cr}=5819$, $Re_{cr}=50$, $m_{cr}=0$.

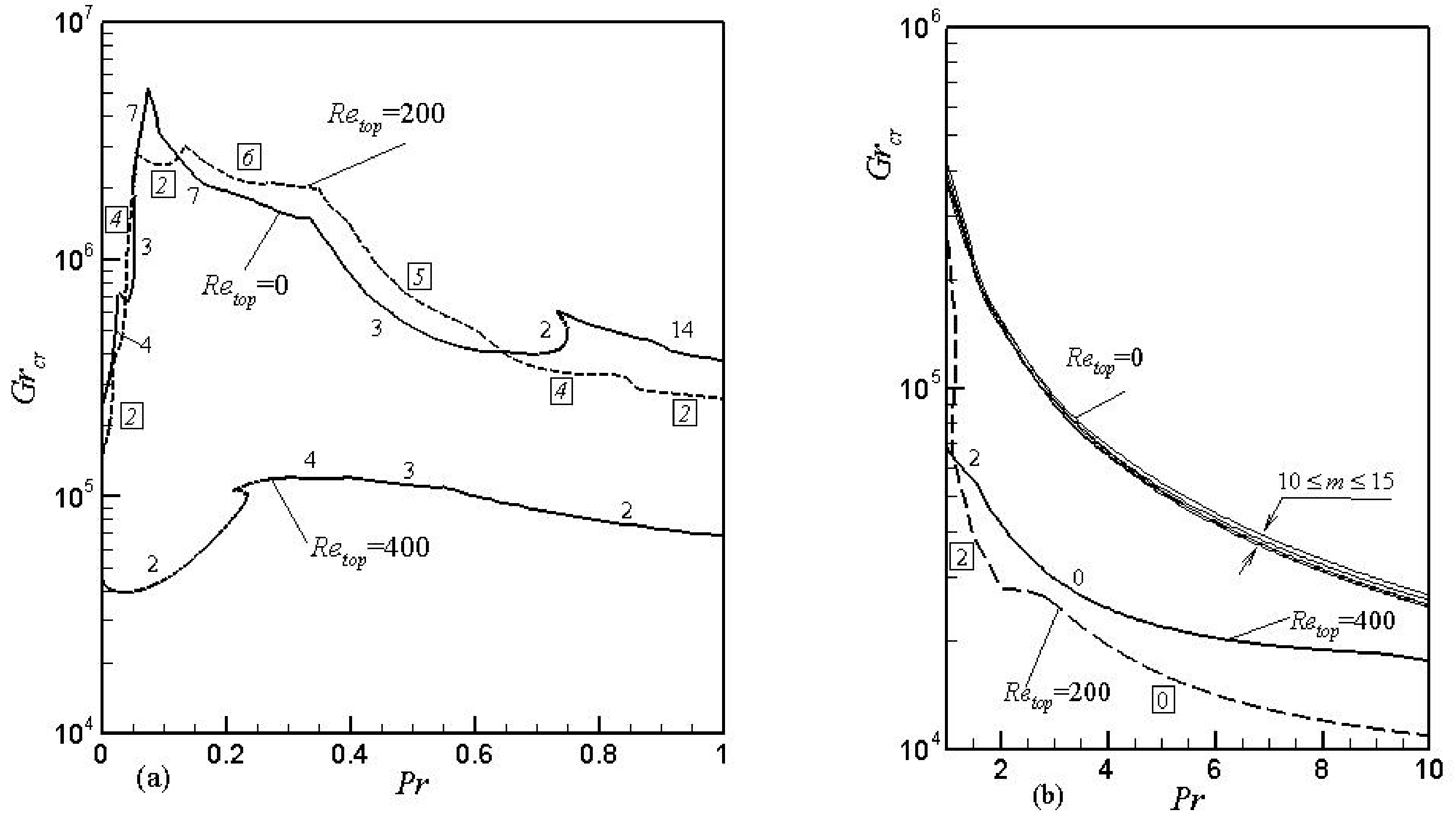

Figure 14. Neutral stability curves $Gr_{cr}(Pr)$ for flow in a vertical cylinder with parabolic temperature profile at the sidewall and rotating top for fixed values of the Reynolds number $Re$=0, 200, and 400. Numbers on the curves correspond to the critical Fourier mode $m_{cr}$. The neutral curve for $Re$=200 is shown by dashed lines and the corresponding numbers are framed.

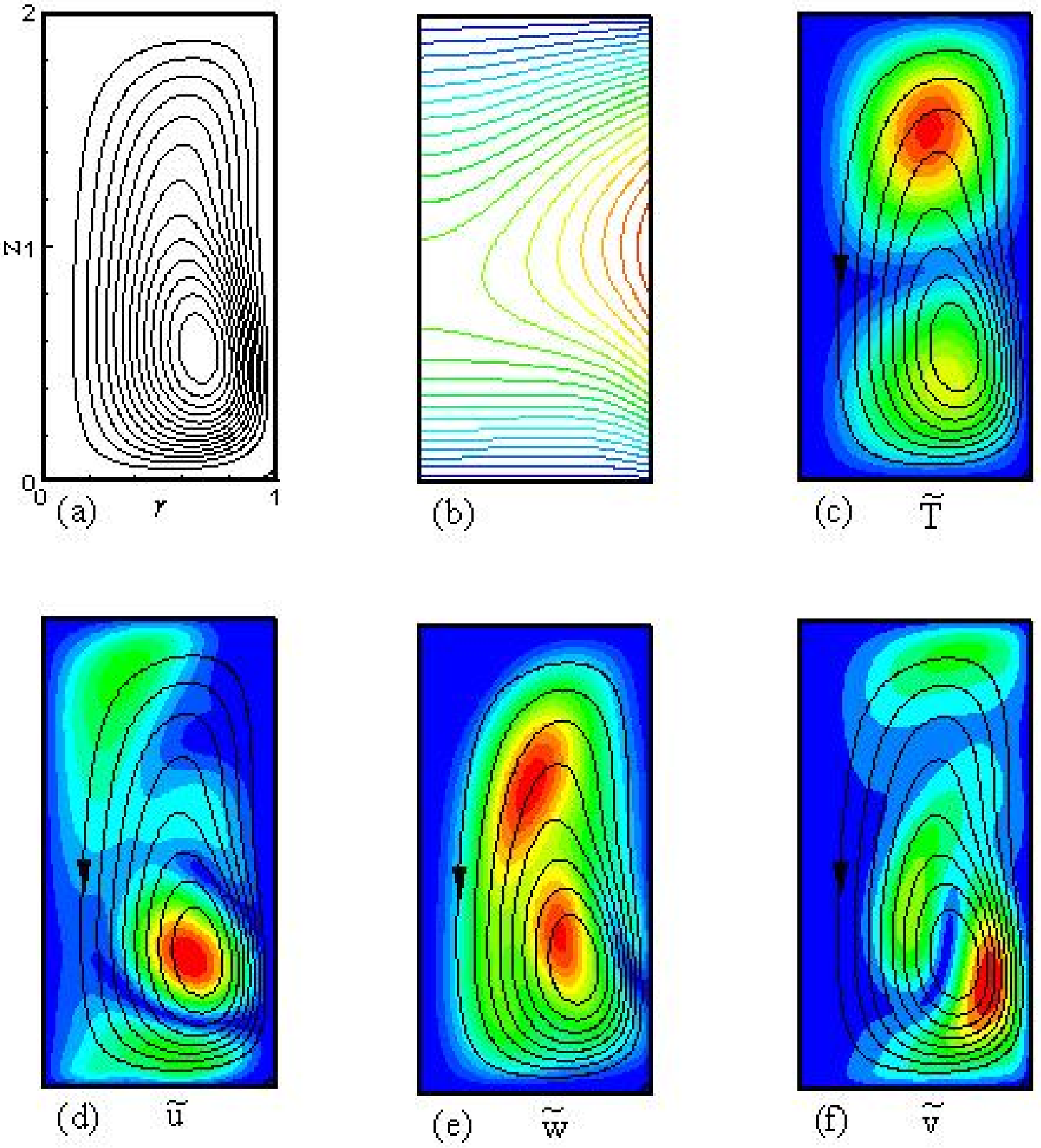

Fig. 15. Flow and perturbation patterns for a purely convective flow with a parabolic heating at the sidewall. $A$=2, $Pr$=0.015, $Gr_{cr}$=1.01×10$^5$. Instability for the azimuthal wavenumber $m$=2. (a) streamlines and (b) isotherms of the flow. Streamlines are evenly distributed between the values of 0 and -27 (global minimum). Isotherms are evenly distributed between the values 0 and 1. Frames (c), (d), (e) and (f) show absolute value of the most unstable perturbation of the temperature and radial, axial and azimuthal velocities, respectively.

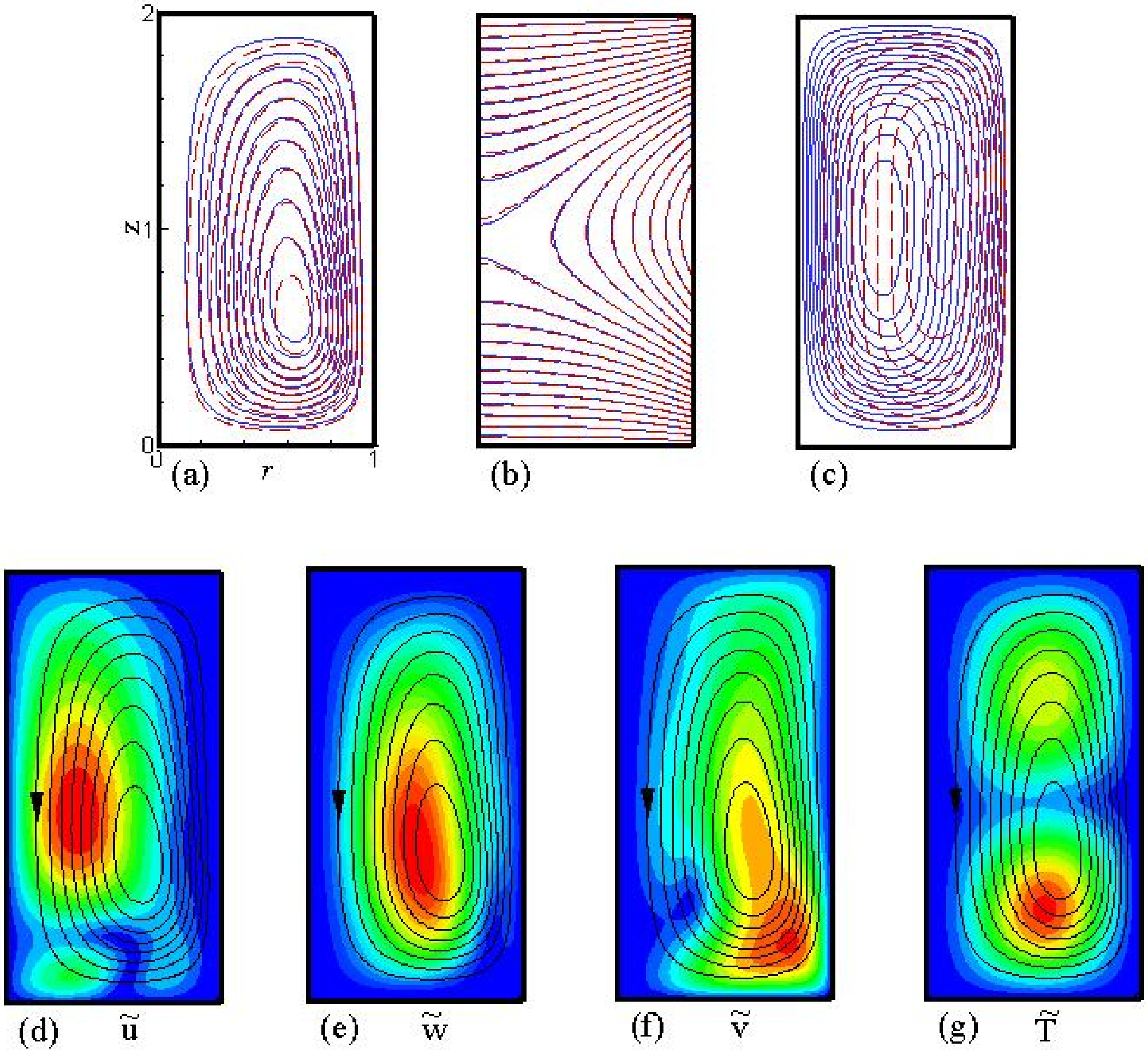

Fig. 16. Flow and perturbation patterns for a purely convective flow with a parabolic heating at the sidewall under effect of a rotating magnetic field. $A$=2, $Pr$=0.015, $Gr_{\mathrm{cr}}=2\times10^4$, $Ta$=1020. Instability for the azimuthal wavenumber $m$=2. (a) streamlines and (b) isotherms of the considered flow (blue solid line) and of the flow at $Ta$=0 at the same $Gr$ (red dash line). (c) Isolines of the azimuthal velocity of the considered flow (blue solid line) and of the RMF-driven flow with $Gr$=0. Streamlines are evenly distributed between the values of 0 and –8 (global minimum), and isolines of the azimuthal velocity between 0 and 40. Isotherms are evenly distributed between the values 0 and 1. Frames (d), (e), (f) and (g) show absolute value of the most unstable perturbation of the radial, axial and azimuthal velocities, and the temperature, respectively.

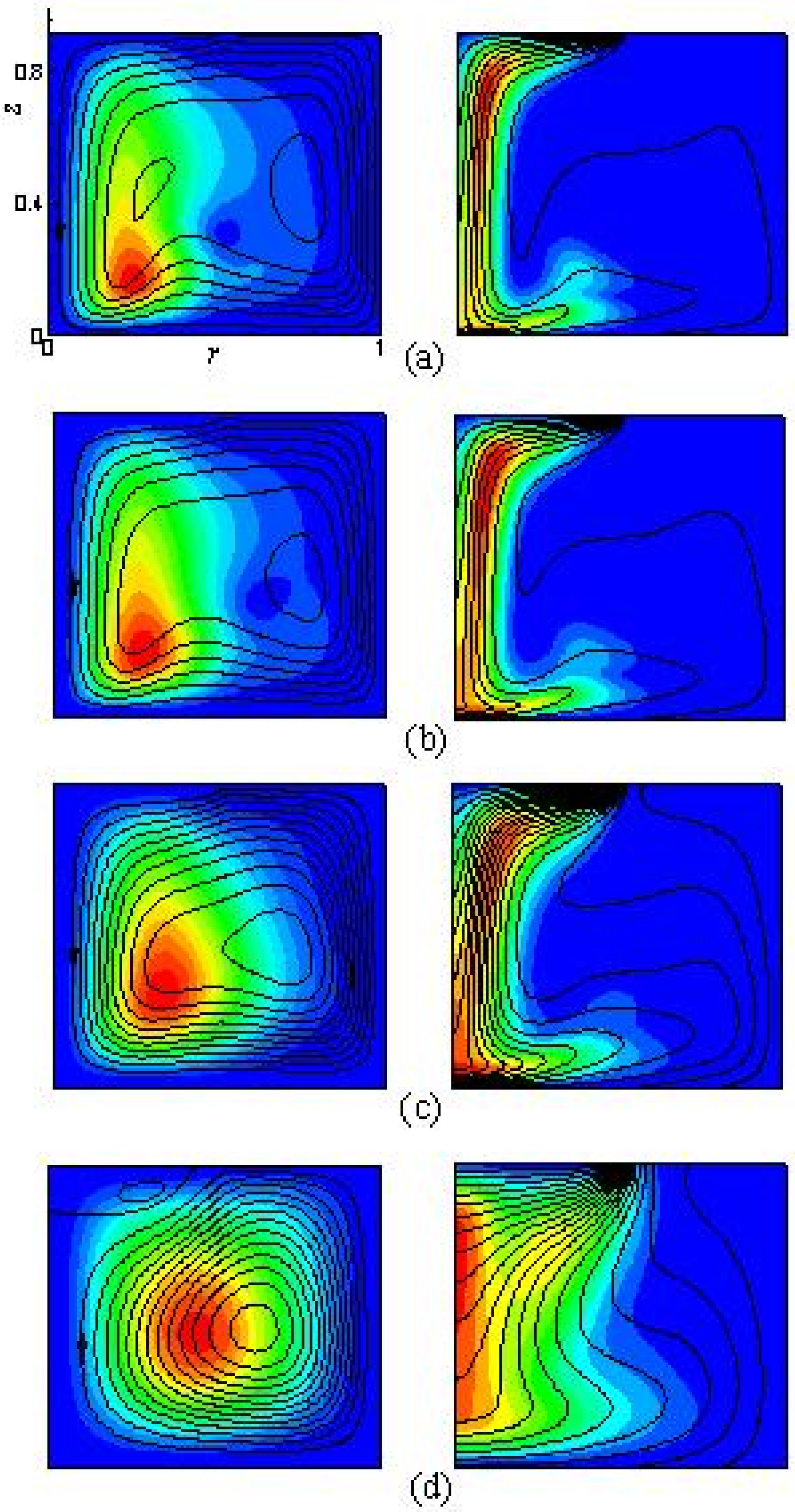

Fig. 17. Streamlines (left frames) and isotherms (right frames) shown by lines and perturbations of the stream function and temperature shown by color for four points a, b, c, and d shown in Fig. 1a. $m_{cr}$=0. (a) $\Delta T$=0.63, $Re$=0, $\psi_{min}$=–1.38, (b) $\Delta T$=0.37, $Re$=75, $\psi_{min}$=–1.22, (c) $\Delta T$=0.1, $Re$=104, $\psi_{min}$=–0.81, (d) $\Delta T$=0.018, $Re$=100, $\psi_{min}$=–0.40, $\psi_{max}$=–0.0066.

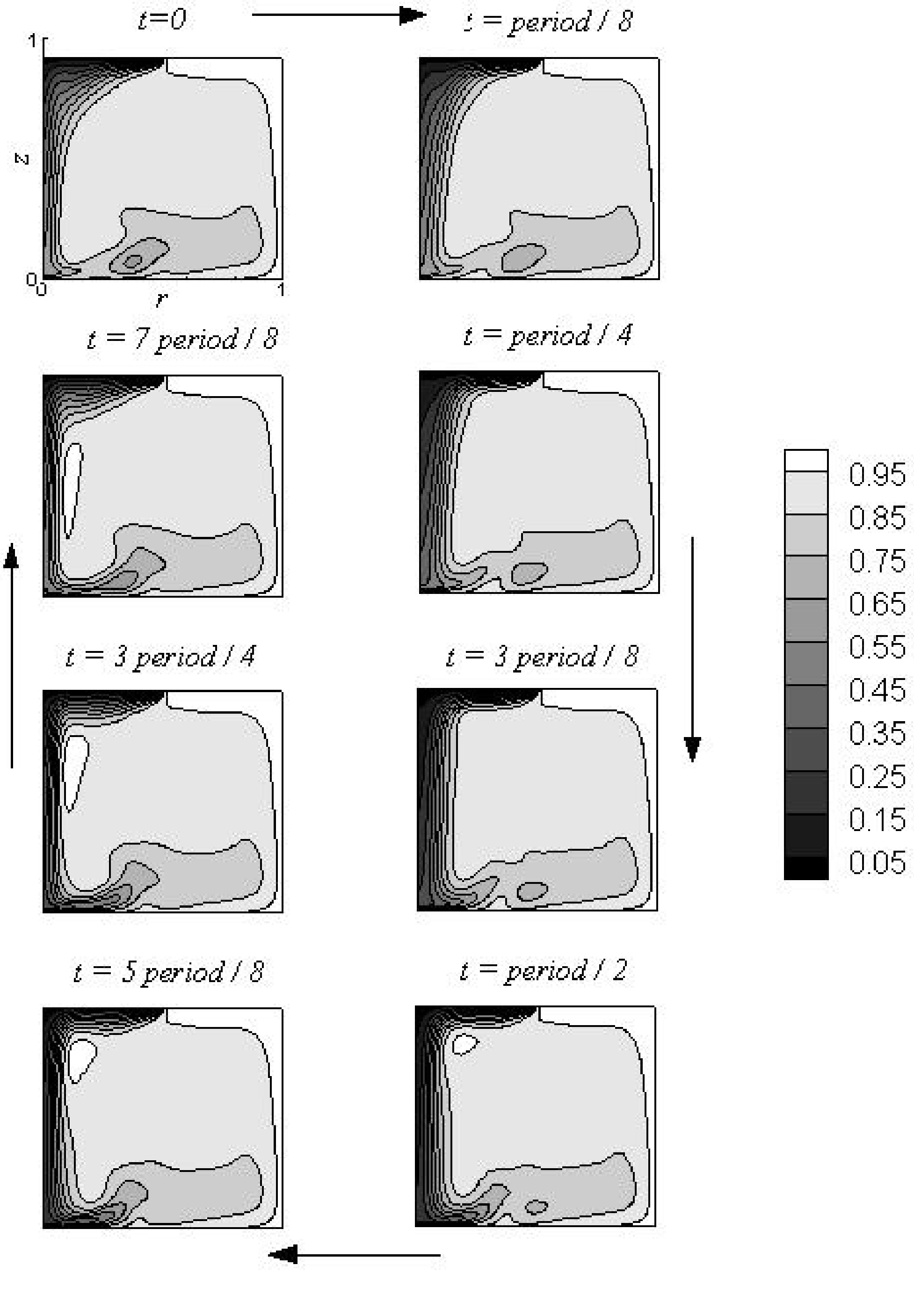

Fig. 18. Eight equally distanced snapshots of isotherms of supercritical oscillatory state during one period of oscillations. Flow in the Czochralski model of Crnogorac et al. (2008) top (Fig. 1a, point a) at $\Delta T=0.63$, $Re_{cr}=0$, $m_{cr}=0$.

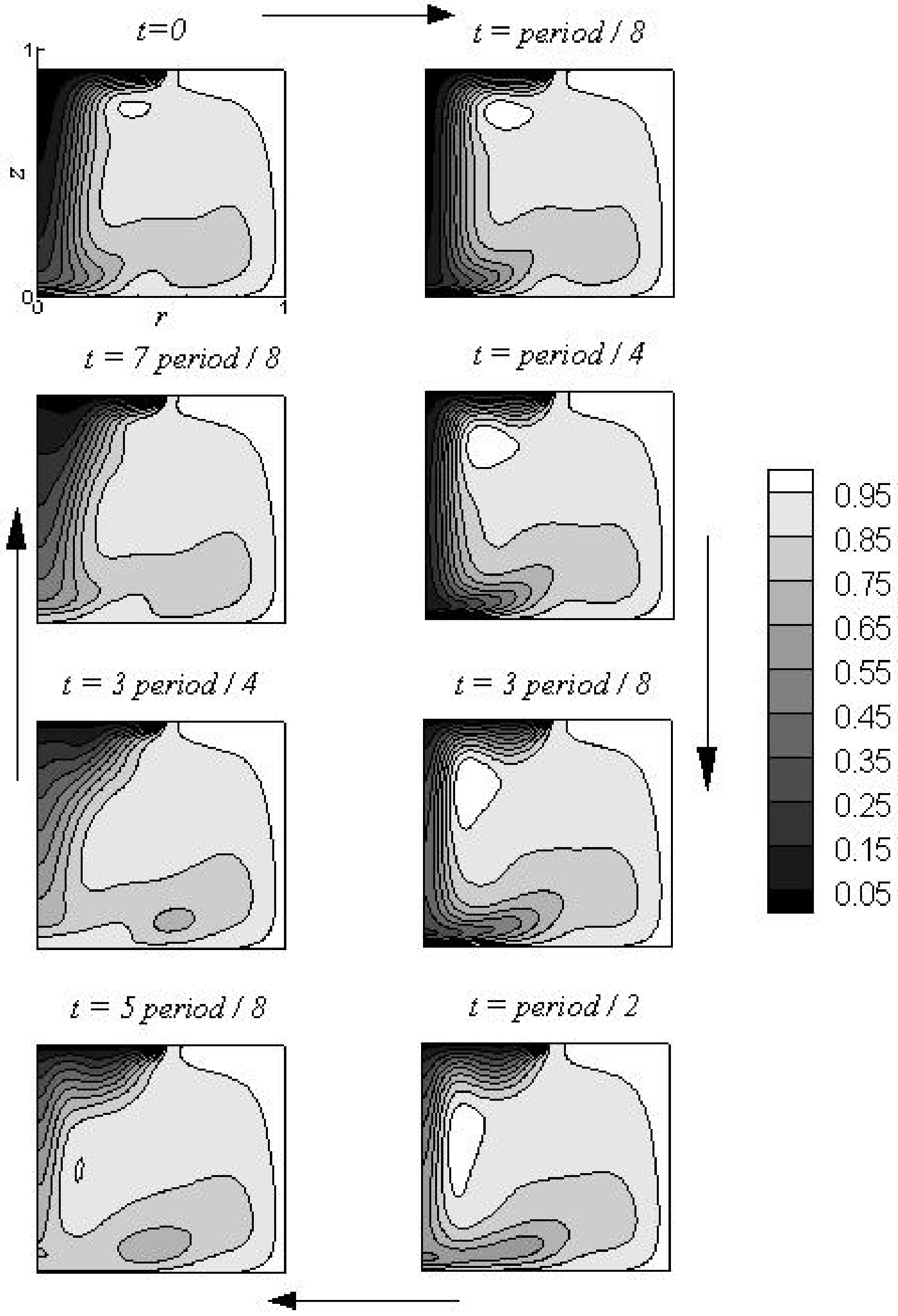

Fig. 19. Eight equally distanced snapshots of isotherms of supercritical oscillatory state during one period of oscillations. Flow in the Czochralski model of Crnogorac et al. (2008) top (Fig. 1a, point a) at $\Delta T$=0.1, $Re_{cr}=100$, $m_{cr}=0$.